\documentclass[aps, a4paper, amsfonts, amssymb, amsmath, reprint, showkeys, nofootinbib, twoside, floatfix, physrev]{revtex4-2}
\usepackage[english]{babel}
\usepackage[utf8]{inputenc}
\usepackage{wrapfig}
\usepackage{float}
\usepackage[colorinlistoftodos, color=green!40, prependcaption]{todonotes}
\usepackage{comment}
\usepackage{amsthm}
\usepackage{mathtools}
\usepackage{physics}
\usepackage{xcolor}
\usepackage{graphicx}
\usepackage[left=23mm,right=13mm,top=35mm,columnsep=15pt]{geometry} 
\usepackage{adjustbox}
\usepackage{placeins}
\usepackage[T1]{fontenc}
\usepackage{lipsum}
\usepackage{csquotes}
\usepackage[pdftex, pdftitle={Article}, pdfauthor={Author}]{hyperref} 

\newcommand{\uprightsuperscript}[1]{^{\textnormal{#1}}}
\begingroup\lccode`~=`?\lowercase{\endgroup\let~}\uprightsuperscript
\AtBeginDocument{\mathcode`?="8000 }

\begin{document}
\title{High gain squeezing in lossy resonators:  an asymptotic field approach
}

\author{M. Sloan$^1$, A. Viola$^2$, M. Liscidini$^2$, and J. E. Sipe $^1$}
\affiliation{$^1$Department of Physics, University of Toronto, 60 St, George Street, Toronto, ON, M5S 1A7, Canada}
\affiliation{$^2$Dipartimento di Fisica, Università di Pavia, Via Bassi 6, 27100 Pavia, Italy}

\date{\today} 

\begin{abstract}
We present a method for describing nonlinear electromagnetic interactions in integrated photonic devices utilizing an asymptotic-in/out field formalism. Our method expands upon previous continuous wave asymptotic treatments by describing the evolution non-perturbatively for an arbitrary pulsed input. This is presented in the context of a squeezing interaction within an integrated microring resonator side coupled to an input/output waveguide, but is readily generalizable to other integrated structures, while including a variety of (non-squeezing) third-order interactions. An example of a single-pump, non-degenerate squeezing interaction is studied, which is shown to match well with standard coupled-mode treatments for high-finesse resonators, as well as previous perturbative treatments dealing with the generation of pairs with low probability.
\end{abstract}

\maketitle

\section{Introduction} \label{sec:Introduction}
Nonlinear optical interactions in integrated photonic devices enable the generation and manipulation of a wide range of exotic states of light. An important example is squeezed light, where the noise in one quadrature component is reduced below the vacuum level; it has seen a broad range of applications in metrology \cite{Caves-PhysRevD.23.1693, LIGO} and imaging \cite{Kolobov:93, Imaging, Ttreps-PhysRevLett.88.203601}, and is a necessary resource for many photonic quantum computing strategies \cite{Bourassa2021blueprintscalable, Filip-PhysRevA.71.042308}. Additionally, through the use of linear optical elements and photon number resolving detectors, a variety of non-classical states of light, such as NOON states \cite{Boto-PhysRevLett.85.2733}, cat states \cite{Gerry-10.1119/1.18698}, GKP states \cite{Gottesman-PhysRevA.64.012310}, and W-states \cite{Dur-PhysRevA.62.062314} can be generated from a squeezed input \cite{Su-PhysRevA.100.052301, Quesada-PhysRevA.100.022341, Gerrits-PhysRevA.82.031802}.

Squeezed light can be generated utilizing nonlinear optical processes such as spontaneous four-wave mixing (SFWM) \cite{Slusher:87, Heersink:05} and spontaneous parametric down conversion (SPDC) \cite{Pereira-PhysRevA.38.4931, Mehmet:11, Mehmet_2019} in materials with an appreciable nonlinear permittivity. Advances in fabrication have allowed for the construction of integrated micro-cavities that benefit from increased scalability compared to bulk crystals, with many optical components integrated on a single chip \cite{Bogaerts}. Indeed, squeezed light generation has been demonstrated in a variety of integrated platforms, including periodically poled waveguides \cite{Lenzini-doi:10.1126/sciadv.aat9331, Mondain:19, Kashiwazaki-10.1063/5.0144385}, microring resonators \cite{Dutt-PhysRevApplied.3.044005, Vaidya-doi:10.1126/sciadv.aba9186, Zhao-PhysRevLett.124.193601}, and multi-ring ``photonic molecules'' \cite{Zhang_2021}. 
    
In particular, microring resonators are an attractive structure for squeezing, as they allow for large field enhancements, with the generated fields being restricted to a discrete set of resonances. However, many theoretical treatments of nonlinear pair generation in these devices typically model the squeezing as occurring within modes corresponding to an isolated ring \cite{Quesada:22, Seifoory, Helt:10, Vernon-PhysRevA.92.033840, Cui-PhysRevResearch.3.013199}, with the fields entering and exiting through an input/output waveguide coupled to the ring at a point; this is a so-called ``coupled mode'' approach. Alternately, treatments of the nonlinear field interaction of a coupled waveguide and ring have been explored using an asymptotic-in/out formalism \cite{Breit, Liscidini, Banic-PhysRevA.106.043707} in which the field modes are defined over the whole coupled system. While a standard coupled mode treatment necessarily assumes a high finesse ring with Lorentzian field enhancement, asymptotic field methods can be applied even in low finesse and allow for a general description of the coupling between the ring and the waveguide, as well as of the variation of the field profile within the resonator structure. This can become particularly important for systems with a spatially varying coupling, such as those utilizing a pulley configuration \cite{Moille:19}, or in cases where a description of the nonlinear interaction within the coupling region is required, such as in dual ring photonic molecules with linearly uncoupled rings \cite{Tan:20, Menotti}. Despite this, to date asymptotic field treatments have only been applied to perturbative calculations \cite{Liscidini, Banic-PhysRevA.106.043707} that allow for the computation of the photon generation rate for sufficiently weak pump powers, but do not reconstruct the operator moments of the generated state needed for determining the squeezing and broadband coherence values \cite{Quesada:22, Christ_2011}. 

In this work, we develop a method for describing the evolution of the pump and generated fields in a coupled ring and waveguide system utilizing an asymptotic field expansion of the mode basis. Such a method extends previous asymptotic treatments by describing the evolution non-perturbatively, while including parasitic third order non-squeezing interactions. Additionally, we perform this calculation while allowing for the coupling between the ring and the input/output waveguide to occur over a directional coupler of finite length. While we use the example of a single ring coupled to a single waveguide as an example to illustrate our general approach, we emphasize that this treatment can be generalized to other resonant systems, including microtoroids \cite{Zhang:13}, whispering gallery resonators \cite{Lin:17}, and multi-ring resonators \cite{Zhang_2021}, while allowing for an arbitrary description of the resonator and waveguide coupling.

In Sec \ref{sec:AsymptoticFields} and \ref{sec:LocalBasis} we introduce the structure of interest and develop the asymptotic-in/out modes of the system. A local basis description of the fields inside the resonator is constructed, which is used to simplify the following derivation. In Sec \ref{sec:NonlinearInteraction} and Sec \ref{sec:SampleCalculation} we develop the equations of motion for the relevant operators, and apply the treatment to a single pump, non-degenerate SFWM squeezing interaction. From this, we calculate the second order moments of the bosonic creation and annihilation operators and relate these quantities to the pair generation rate, broadband correlation values, and squeezing spectrum of the output fields as outlined in Sec \ref{sec:Results}. Finally, we conclude with Sec \ref{sec:Conclusion}.

\section{Asymptotic Fields} \label{sec:AsymptoticFields}
We begin with the asymptotic fields for a single lossless ring, side coupled to a waveguide along a finite coupling length; the strategy for modeling this simple system will serve as a guide to the more complicated configurations that will follow.

\subsection{Lossless System}
\begin{figure}
    \centering
    \includegraphics[width=0.45\textwidth]{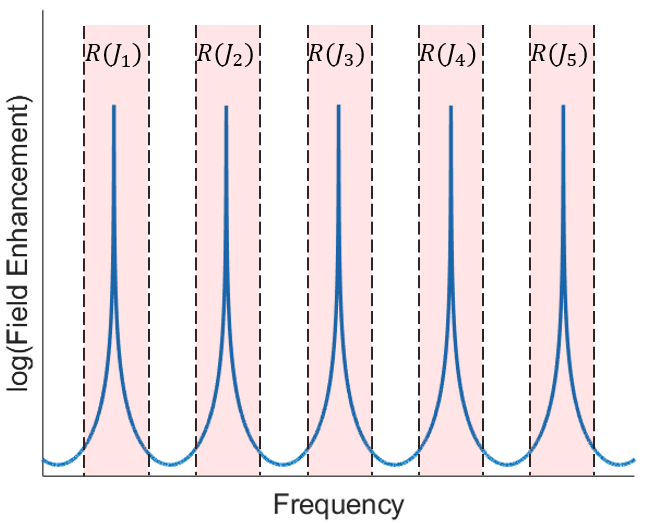}
    \caption{Diagram of the field enhancement inside a single ring resonator coupled to an input/output waveguide. For each resonance $J_i$ we define a range $R(J_i)$ which contains the resonance, but is disjoint from each $R(J_j)$ for $j \neq i$.}
    \label{fig:GeneralEnhancementPlot_withBins}
\end{figure}

To start, we consider an infinitely long waveguide extending along the $z$ direction; we allow the index of refraction to vary with the $x$ and $y$ coordinates in such a way that the optical modes of interest are confined to the waveguide, but we assume that the structure is uniform along the $z$ axis. In what is to follow we will be interested in such a waveguide coupled to a ring resonator, with a pump field having $k-$space support near one ring resonance, and generated fields associated with one or more of the resonances. As such, we let $J$ index the resonances of the coupled system and introduce the frequency ranges $R(J)$ which contain the resonance $J$ but are disjoint from all $R(J')$ whenever $J' \neq J$ (See Fig. (\ref{fig:GeneralEnhancementPlot_withBins})). Choosing $k_J$ to be a reference wavevector in the range $R(J)$, this allows us to write the displacement field for the isolated waveguide as \cite{Quesada:22}
\begin{equation} \label{eq:ChannelFieldDisplacementOperator}
\begin{split}
    \textbf{D}(\textbf{r}, t) &= \sum_{J} \int_{R(J)} dk \sqrt{\frac{\hbar \omega_{k}}{4 \pi}} \textbf{d}?{wg}_{k}(x,y) a(k,t) e^{ikz} + H.c. \\
    & \cong \sum_{J} \sqrt{\frac{\hbar \omega_{J}}{2}} \textbf{d}?{wg}_{J}(x,y) \psi_{J}(z, t)e^{ik_J z} + H.c.,
\end{split}
\end{equation}
where $a(k,t)$ denotes the Heisenberg picture annihilation operator corresponding a wavevector $k$, with associated frequency $\omega_{k}$, and the operators $\psi_J(z,t)$ are defined as
\begin{equation}
    \psi_{J}(z,t) \equiv \frac{1}{\sqrt{2\pi}} \int_{R(J)} a(k,t) e^{i(k - k_J)z} dk.
\end{equation}
The functions $\textbf{d}_k?{wg}(x,y)$ describes the field profile along the plane perpendicular to the waveguide length. We have made the assumption that the ranges $R(J)$ are small enough that we can set $\omega_{k} \cong \omega_{J} \equiv \omega_{k_J}$ in the square root term in (\ref{eq:ChannelFieldDisplacementOperator}), as well as $\textbf{d}_k?{wg}(x,y) \cong \textbf{d}_J?{wg}(x,y) \equiv \textbf{d}_{k_J}?{wg}(x,y)$.

For the $a(k,t)$ and $a^{\dagger}(k,t)$ to satisfy the canonical commutation relations, 
\begin{equation} \label{eq:bosonicCommutationRelation}
  [a(k,t),a^{\dagger}(k',t)] = \delta(k-k'),
\end{equation}  
the field distributions $\textbf{d}?{wg}_{J}(x,y)$ should be normalized according to
\begin{equation}
    \int \frac{\textbf{d}?{wg*}_{J}(x,y) \cdot \textbf{d}?{wg}_{J}(x,y)}{\epsilon_0\epsilon(x,y)} dx dy = 1,
\end{equation}
for a relative dielectric permittivity $\epsilon(x,y)$, where here we neglect any frequency dependence of 
the dielectric constant over the range R(J); however, this can be generalized to include such a variation if necessary \cite{Bhat-PhysRevA.73.063808, Quesada:22}. For simplicity, we consider the input pump field, as well as any generated fields, to be in one spatial mode of the system with polarization perpendicular to the plane of the ring and waveguide, but this can also easily be generalized.

With the above expansion in mind, if we were to neglect any nonlinearity the operators $a(k,t)$ would evolve according to
\begin{equation} \label{eq:LinearEvolutionKoperator}
    a(k,t) = a(k, 0)e^{-i \omega_{k} t}.
\end{equation}
Assuming group velocity dispersion is negligible within each range $R(J)$, for any $k$ in $R(J)$ we can write 
\begin{equation} \label{eq:waveguideDispersionRelation}
    \omega_{k} = \omega_{J} + v_{J}(k - k_J),
\end{equation}
with $v_{J}$ the group velocity within that range;
it immediately follows that the operator $\psi_J(z,t)$ satisfies the equation of motion
\begin{equation} \label{eq:LinearChannelFieldEOM}
\begin{split}
    &\frac{\partial}{\partial t} \psi_{J}(z,t) = -i\omega_{J} \psi_{J}(z,t) - v_{J} \frac{\partial}{\partial z} \psi_{J}(z,t),
\end{split}
\end{equation}
in the absence of nonlinearities.
\begin{figure*}
    \centering
    \includegraphics[width=\textwidth]{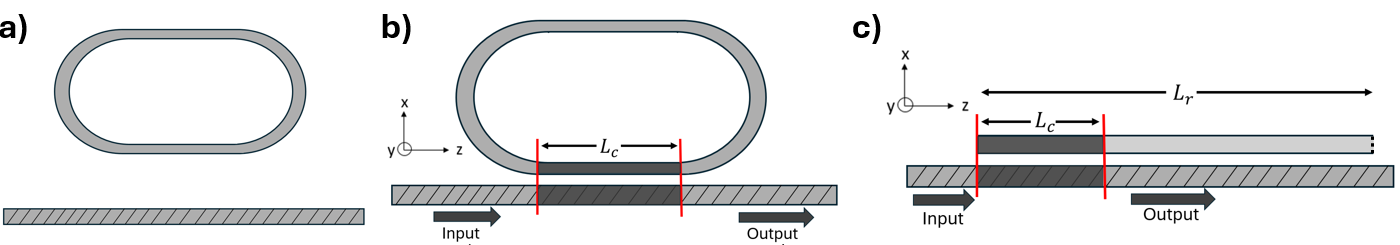}
    \caption{(a) Isolated ring resonator and channel waveguide. (b) Diagram of the lossless ring and waveguide system. Coupling between the structure is allowed to occur over a finite length denoted by the shaded region. (c) Simplified waveguide and resonator system in which the ring is approximated as a separate waveguide of finite length. The two ends of the finite waveguide are identified as the same point.}
    \label{fig:LosslessSystem}
\end{figure*}

Next we consider an isolated racetrack resonator (see Fig. \ref{fig:LosslessSystem}a) and decompose the displacement operator in the resonator into the distinct frequency bins introduced above. In general, the mode expansion for this type of structure will be more complicated than that for the straight waveguide due to, for example, changes in the field distribution along the cross section of the resonator as one moves from the straight segment to the curved section. However, if we first restrict ourselves to the straight region of the resonator, then we can expand the ring field locally in the same way as it was done for the waveguide. Orienting the straight region along the $z$-axis and letting $x$ and $y$ be the "local" coordinates running perpendicular to the direction of propagation, we can write the displacement field in this region as
\begin{equation} \label{eq:ringFieldExpansion}
    \textbf{D}(\textbf{r}, t) = \sum_{J} \sqrt{\frac{\hbar \omega_{J}}{2}} \textbf{d}?{rr}_{J}(x,y) \phi_{J}(z, t)e^{ik_J' z} + H.c.,
\end{equation}
with $\phi_J(z,t)$ the analogue of the operator $\psi_J(z,t)$ in the waveguide and $\textbf{d}?{rr}_J(x,y)$ describing the field profile along the plane tangent to the propagation direction. For the fields within the ring structure, we expand the dispersion relation around $k_J'$ as $\omega_{k} = \omega_{J} + u_{J}\left(k - k_J' \right)$. Here for any given range $R(J)$ we have chosen the reference $\omega_{J}$ to be the same for both the waveguide and the ring, but we allow the group velocity in the ring, $u_{J}$, as well as the reference wavenumber in the ring, $k_J'$, to differ from their values in the waveguide. In particular, we will choose $k_J$ and $k_J'$ such that the angular frequency $\omega_J$ corresponds to a resonance peak of the isolated ring.

With this, it follows that the equations of motion for the operators $\phi_{J}(z,t)$ are locally given by
\begin{equation} \label{eq:uncoupledLinearEOMring}
    \frac{\partial}{\partial t} \phi_{J}(z,t) = -i\omega_{J} \phi_{J}(z,t) - u_{J} \frac{\partial}{\partial z} \phi_{J}(z,t).
\end{equation}

From the expansion above, one could then expand the description to the full ring structure by solving for the field distribution outside the straight sections, and relating it back to the fields at the boundaries. However, in the case when the bending radii of the curved regions of the resonator are taken to be large, one would expect the field distribution along the plane perpendicular to the direction of propagation to vary little from 
$\textbf{d}?{rr}_J(x,y)$ for the straight section. As such, one could approximate the resonator as being equivalent to a straight waveguide of effective length $L_r$, with endpoints identified together, and extend the expansion of the displacement operator in equation (\ref{eq:ringFieldExpansion}) to the full effective length of the resonator. For a first calculation, we will adopt this perspective and represent the coupled ring and waveguide system in Fig. \ref{fig:LosslessSystem}b as equivalent to scheme shown in Fig. \ref{fig:LosslessSystem}c. This approximation will be convenient as it will allow us to determine the asymptotic field distributions analytically. But we emphasize that one could extend this to a general resonator structure, with the variation in the waveguide curvature leading to spatial variations in the effective index, group velocity, and scattering loss, as well as allowing for coupling to higher order spatial modes of the resonator.

With the displacement operators of the isolated waveguides and ring resonator in hand, we can now consider bringing the structures close enough that they can share energy through their overlapping evanescent fields, but sufficiently far away that the field transfer can be approximated through the addition of linear coupling terms between the $\psi_J(z,t)$ and $\phi_J(z,t)$ operators in equations (\ref{eq:LinearChannelFieldEOM}) and (\ref{eq:uncoupledLinearEOMring}) \cite{microRingBook}. Choosing for both structures to be oriented along the $z$ axis, and allowing for the ring and waveguide coupling to occur over the range $[0, L_c]$, we can place one end of the now 'straight' resonator at $z=0$, and introduce a coupling rate $\omega^c_J$ such that the equations of motion for the coupled system is given by \cite{microRingBook}
\begin{widetext}
\begin{equation} \label{eq:EOMofLinearFields}
\begin{split}
    \frac{\partial}{\partial t} \psi_{J}(z,t) &= 
    \begin{cases}
        -i\omega_{J} \psi_{J}(z,t) - v_{J} \frac{\partial}{\partial z} \psi_{J}(z,t) - i\omega_J^c \phi_J(z,t) e^{-i(k_J - k_J')z} & \textit{for } 0 < z < L_c \\
        -i\omega_{J} \psi_{J}(z,t) - v_{J} \frac{\partial}{\partial z} \psi_{J}(z,t) & \textit{otherwise}
    \end{cases} \\
    \frac{\partial}{\partial t} \phi_{J}(z,t) &= 
    \begin{cases}
        -i\omega_{J} \phi_{J}(z,t) - u_{J} \frac{\partial}{\partial z} \phi_{J}(z,t) - i\omega_J^c \psi_J(z,t)e^{i(k_J - k_J')z} & \textit{for } 0 < z < L_c \\
        -i\omega_{J} \phi_{J}(z,t) - u_{J} \frac{\partial}{\partial z} \phi_{J}(z,t) & \textit{for } L_c < z < L_r
    \end{cases}.
\end{split}
\end{equation}
\end{widetext}
Here the phase in the last term of the equations of motion within the coupling region takes into account the phase mismatch between the waveguide and the ring fields.

At this point, we have developed a description of the displacement fields in terms of the fields for each isolated segments of the system. However, we would like to use the local expansion of the fields in terms of $\psi_J(z,t)$ and $\phi_J(z,t)$ to develop an expansion of the displacement field in terms of the asymptotic-in(out) mode fields \cite{Breit, Liscidini, Banic-PhysRevA.106.043707}. With this approach we can represent our fields by modes defined over all space, with the transfer of energy between elements taken into account through the construction of the field distributions of the modes, rather than through additional coupling terms in the system Hamiltonian.

In particular, we will write the displacement field for the full coupled system as \cite{Breit, Liscidini}
\begin{equation} \label{eq:LosslessAsyFields}
\begin{split}
    \textbf{D}(\textbf{r}, t) &= \sum_J \int_{R(J)} dk \sqrt{\frac{\hbar \omega_J}{4\pi}} \textbf{D}?{asy-in(out)}_{J,k}(\textbf{r}) \\
    & \qquad \qquad \qquad \qquad \times a?{in(out)}_J(k,t) + H.c.,
\end{split}
\end{equation}
where $\textbf{D}?{asy-in(out)}_{J,k}(\textbf{r})$ describes the field for a monochromatic wave input (output) from the channel waveguide and $a?{in(out)}_J(k,t)$ is the associated annihilation operator. Upon introducing a parameter $\tau = \tau(\textbf{r}) \in \{0, 1\}$ that takes the value 0 when $\textbf{r}$ is located in the waveguide and 1 when $\textbf{r}$ is in the resonator, it can be decomposed within the material region as
\begin{equation} \label{eq:fieldDistDecom}
    \textbf{D}?{asy-in(out)}_{J,k}(\textbf{r}) = \textbf{d}?{asy}_{J, \tau}(x,y) h?{in(out)}_{J,k, \tau}(z) e^{ik_J' z}.
\end{equation}

If one were to place the system such that the center point of the waveguide cross section is located at $(x,y) = (0,0)$, with the center point of the cross section of the ``straightened resonator'' at $(x,y) = (x_0, 0)$, then we could write
\begin{equation}
    \textbf{d}?{asy}_{J, \tau}(x,y) =
    \begin{cases}
        \textbf{d}?{wg}_{J}(x,y) & \textit{for } \tau=0 \\
        \textbf{d}?{rr}_{J}(x-x_0,y) & \textit{for } \tau = 1 \\
    \end{cases}.
\end{equation}
The other relevant function, $h?{in(out)}_{J,k}(z)$, gives the slowly varying $z$ dependence of the field distribution, which can be related to the classical solutions to the system in equation (\ref{eq:EOMofLinearFields}) in the case of an input (output) in the waveguide of $e^{ikz}$. So making the replacement 
\begin{equation} \label{eq:classicalApproximation}
\begin{split}
    \psi_J(z,t) & \rightarrow \langle \psi_J(z,t) \rangle \\
    \phi_J(z,t) & \rightarrow \langle \phi_J(z,t) \rangle ,
\end{split}
\end{equation}
in equation (\ref{eq:EOMofLinearFields}), we can introduce the function $l_J(z)$ and an effective resonator length $\tilde{L}_J$ such that
\begin{equation}
    l_J(z) \equiv \begin{cases}
        \frac{1}{2} \left( 1 + \frac{v_J}{u_J} \right) z & \textit{for } 0 < z < L_c, \\
        \frac{v_J}{u_J} z + \frac{1}{2} \left( 1 - \frac{v_J}{u_J} \right)L_c & \textit{for } L_c < z < L_r,
    \end{cases}
\end{equation}
and
\begin{equation}
    \tilde{L}_J \equiv l_J(L_r) = \frac{v_J}{u_J} \left[ L_r - \frac{1}{2} \left( 1 - \frac{u_J}{v_J} \right)L_c \right],
\end{equation}
allowing us to write the solutions to the system for a given input $k$ as (see Appendix \ref{sec:appendix1})
\begin{widetext}
\begin{equation} \label{eq:sampleClassicalSolution}
\begin{split}
    \langle \psi_J(z,t; k) \rangle &= \langle \psi_J(0,t; k) \rangle
    e^{-i\omega_k t} \begin{cases}
        e^{i\Delta k_J z} & \textit{for } z < 0 \\
        \left[ \sigma_J(z;k) - i\sqrt{\frac{u_J}{v_J}} R_J(k) \kappa_J(z;k) e^{i \Delta k_J \tilde{L}_J} \right] e^{i \Delta k_J l_J(z)} & \textit{for } 0 < z < L_c \\
        T_J(k) e^{i\Delta k_J (z - L_c)} & \textit{for } z > L_c 
    \end{cases} \\
    \langle \phi_J(z,t; k) \rangle &= \langle \psi_J(0,t; k) \rangle e^{-i\omega_k t} \begin{cases}
         \left[ R_J(k) \sigma_J^*(z;k)e^{i \Delta k_J \tilde{L}_J} - i \sqrt{\frac{v_J}{u_J}} \kappa_J^*(z;k) \right] e^{i \Delta k_J l_J(z)} & \textit{for } 0 < z < L_c \\
        R_J(k) e^{i \Delta k_J l_J(z)} & \textit{for } z > L_c 
    \end{cases},
\end{split}
\end{equation}
\end{widetext}
with $\Delta k = k - k_J$ the detuning of the field from the resonance center. Here the functions $\sigma_J(z;k)$ and $\kappa_J(z;k)$ are defined as
\begin{equation} \label{eq:sigmaKappaDef}
    \begin{split}
        \sigma_J(z;k) &= \left[\cos \left(\alpha_k z \right) + i\gamma_J(k) \sin \left(\alpha_k z \right) \right] e^{-i\Delta \beta_J z/2}, \\
        \kappa_J(z;k) &= \sqrt{1 - \gamma_J(k)^2} \sin \left(\alpha_k z \right) e^{-i \Delta \beta_J z/2},
    \end{split}
\end{equation}
where the wavenumber of the sinusoidal envelope is
\begin{equation} \label{eq:alphaDef}
    \alpha_k \equiv \sqrt{\frac{1}{4}\left( \Delta \beta_J + \frac{v_J - u_J}{u_J} \Delta k_J \right)^2 + \alpha_{J,0}^2},
\end{equation}
with $\Delta \beta = k_J - k_J'$ and $\alpha_{J,0} = \frac{\omega_c}{\sqrt{v_J u_J}}$ the envelope wavenumber in the limit $k_J' \rightarrow k_J$ and $u_J \rightarrow v_J$. The parameter $\gamma_J(k)$ is then related to $\alpha_k$ as
\begin{equation} \label{eq:gammaDefinition}
    \gamma_J(k) =  \frac{\sqrt{\alpha_k^2 - \alpha_{J,0}^2}}{\alpha_k},
\end{equation}
and satisfies $0 \leq \gamma_J(k) \leq 1$, with $\gamma_J(k) \rightarrow 0$ in the limit $u_J \rightarrow v_J$ and $k_J' \rightarrow k_J$.

Finally, we define the effective point self-coupling, $\bar{\sigma}_J(k)$, and cross-coupling, $\bar{\kappa}_J(k)$, parameters as $\bar{\sigma}_J(k) = \sigma(L_c;k)$ and $\bar{\kappa}_J(k) = \kappa(L_c;k)$, from which we find
\begin{equation}
    \begin{split}
        R_J(k) &= -i \sqrt{\frac{v_J}{u_J}} \frac{\bar{\kappa}^*_J(k)}{1 - \bar{\sigma}^*_J(k) e^{i\Delta k_J \tilde{L}_J}} \\
        T_J(k) &= \frac{\bar{\sigma}_J(k) - e^{i \Delta k_J \tilde{L}_J}}{ 1 -\bar{\sigma}^*_J(k)e^{i \Delta k_J \tilde{L}_J}} e^{i \Delta k_J l_J(L_c)}
    \end{split}
\end{equation}
Note that for a fixed length of the coupling region there can exist multiple values of the envelope wavenumber $\alpha_k$ that result in the same effective $\bar{\sigma}_J(k)$ and $\bar{\kappa}_J(k)$; these identify the number of times the ring and waveguide field oscillate between the photonic elements within the coupling region.

With equation (\ref{eq:sampleClassicalSolution}) in hand, it then follows that
\begin{equation}
\begin{split}
    & \langle \psi_J(0,0; k) \rangle h?{in}_{J,k,\tau}(z) \\
    & \qquad = \begin{cases}
        \langle \psi_J(z,0; k) \rangle e^{i\Delta \beta_J z} & \textit{for } \tau=0 \\
        \langle \phi_J(z,0; k) \rangle & \textit{for } \tau=1 \textit{, }\\
        & \qquad 0\leq z < L_r \\
    \end{cases}.
\end{split}
\end{equation}
In a similar manner we can find 
$h?{out}_{J,k,\tau}(z)$, choosing the boundary conditions such that $\langle \psi_J(z,0; k) \rangle e^{ik_J' z} = \langle \psi_J(L_c,0; k) \rangle e^{ikz}$ for $z>L_c$ within the waveguide. It should be emphasized here that the field distributions $h?{in}_{J,k,\tau}(z)$ and $h?{out}_{J,k,\tau}(z)$ are not independent. Indeed, in this simple case in which there is only a single input channel and a single output channel, one can show that 
\begin{equation}
\begin{split}
    h?{out}_{J,k,\tau}(z) &= \frac{h?{in}_{J,k,\tau}(z)}{h?{in}_{J,k,0}(L_c)} e^{i\Delta k_J L_c} e^{i \Delta \beta_J L_c} \\
    &= \frac{1 - \bar{\sigma}^*_J(k)e^{i\Delta k \tilde{L}_J} }{\bar{\sigma}_J(k) - e^{i\Delta k \tilde{L}_J} } h?{in}_{J,k,\tau}(z) e^{i \Delta k_J (L_c - l_J(L_c))}
\end{split}
\end{equation}
However, as will be seen in the next section, for a system with many input and output channels, the relationship between the asymptotic-in and the asymptotic-out fields becomes more complicated.

\subsection{Lossy System}
We now expand this treatment to a more realistic setting that includes loss for the pump, signal, and idler fields. Here we are interested in driving the ring and waveguide with a pump frequency well below the material band gap, and with powers sufficiently low that two photon absorption can be safely neglected. In terms of scattering loss, one particularly convenient method of modelling the field attenuation around the resonator, which has been previously used in perturbative treatments with an asymptotic field expansion \cite{Banic-PhysRevA.106.043707}, is to extend the wave vector, $k$, into the complex plane. However, this can create problems in a more general, non-perturbative treatment, since it necessarily neglects the evolution of the lost photons, resulting in a non-unitary evolution of the field operators. An alternate approach to model scattering loss in coupled mode treatments is to introduce a fictitious ``phantom channel'' point coupled to the resonator, through which the photons in the ring can escape \cite{Quesada:22, Seifoory, Helt:10, Vernon-PhysRevA.92.033840, Cui-PhysRevResearch.3.013199}. But 
a straightforward implementation of a single phantom channel with the asymptotic field expansion described thus far would model all of the lost photons as exiting the ring at a single point. This is, of course, unphysical and undesirable if one wants to maintain the correlations between the photons exiting through the waveguide and those scattered at different points within the resonator.

\begin{figure}
    \centering
    \includegraphics[width=0.48\textwidth]{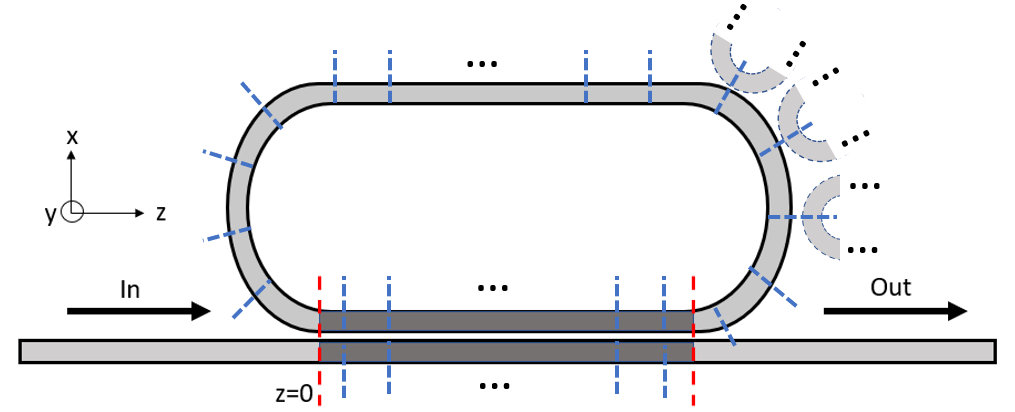}
    \caption{Diagram of the lossy ring and waveguide scenario. Each of the dotted blue lines denotes the coupling to a unique point coupled phantom channel. As in the loss less case, the coupling between the resonator and the waveguide is allowed to occur along the grey shaded region.}
    \label{fig:LossyStructure}
\end{figure}

Instead, we model the scattering loss in the system through the use a number of phantom channels placed along the length of the resonator, as well as in the coupling region of the waveguide (see Fig. \ref{fig:LossyStructure}). As such, if one were to take the number of phantom channels to be very large, then one would recover the exponential amplitude attenuation, while still allowing one to describe the evolution of the lost fields. 

In what is to follow, we denote by $N_L$ the total number of phantom channels coupled to the system. Additionally, for $1 \leq n \leq N_L$, we let $\bar{z}_n$ denote the $z$ position of the coupling point of the $n^{th}$ phantom channel along the waveguide or the ``straightened'' resonator, with $\sigma?{ph}_{J,n}$ and $\kappa_{J,n}?{ph}$ corresponding to the effective self- and cross-coupling parameters. Finally, for each of the phantom channels, we choose the local coordinates such that the length of the channel extends along the $z$ axis, with the point $z=0$ corresponding to the coupling point with the resonator.

With this in mind, the full description of the displacement field would require including the asymptotic field modes corresponding to the waveguide input (output), as well as asymptotic fields corresponding to each of the phantom channel inputs (outputs). Hence, the asymptotic expansion for the lossy system would be given by
\begin{equation} \label{eq:AsymptoticDisplacementField}
\begin{split}
    \textbf{D}(\textbf{r}, t) &= \sum_{J,n} \int_{R(J)} dk \sqrt{\frac{\hbar \omega_J}{4\pi}} \textbf{D}?{asy-in(out)}_{J,k, n}(\textbf{r}) \\
    & \qquad \qquad \qquad \qquad \times a?{in(out)}_{J,n}(k,t) + H.c.,
\end{split}
\end{equation}
(cf. Eq. \ref{eq:LosslessAsyFields}) where we take $a_{J,0}?{in(out)}(k,t)$ to be the Heisenberg picture annihilation operator corresponding to the asymptotic field mode with an input (output) in the waveguide. Then, for $n>0$, each of the $a_{J,n}?{in(out)}(k,t)$ correspond to the annihilation operators for the modes with input (output) in the $n^{th}$ phantom channel, with the field distribution for the mode with the operator $a?{in(out)}_{J,n}(k,t)$ denoted by $\textbf{D}?{asy-in(out)}_{J,k, n}(\textbf{r})$. Expanding each of the $\textbf{D}?{asy-in(out)}_{J,k, n}(\textbf{r})$ as in equation (\ref{eq:fieldDistDecom}), one can follow the same process as described in the previous section to solve for each of the field distributions.   

We note that in distributing the loss around the length of the resonator, as described above, we have the freedom to tune the scattering loss at different points within the ring through modifying the self-coupling parameters, $\sigma?{ph}_{J,n}$, of the phantom channels. Thus, despite approximating the resonator as being equivalent to a waveguide of finite length, we can indeed model the increased scattering loss along the regions corresponding to the bends in the resonator, as well as other forms of loss coming from the particular form of the ring. However, increasing the number of loss channels coupled to the system correspondingly increases the number of modes, leading to longer run times for numerical simulations. As a result, we look to keep $N_L$ modest but large enough to resolve a gradual attenuation of the field in the resonator.

\section{Local Basis} \label{sec:LocalBasis}
With the field expansion as described in the previous section, it becomes clear why the asymptotic-in(out) mode basis gives a convenient way to interpret the evolution of the fields within the structure. Indeed, the field in each of the input (output) channels are completely described by only a single asymptotic-in(out) mode, where by \emph{mode} we mean the set $\left\{ \left(\textbf{D}?{asy-in(out)}_{J,k,n}(\textbf{r}), a?{in(out)}_{J,n}(k,t) \right) \right\}_{k \in R(J)}$ for a particular choice of $n$ and $J$. However, this description suffers from requiring all modes of the system to describe the field at any point within the ring or the coupling region of the waveguide. This is problematic since any nonlinear terms in the equations of motion of each of the operators $a?{asy-in(out)}_{J,n}(k,t)$ would contain complicated overlap integrals over $\textbf{D}?{asy-in(out)}_{J,k,n}(\textbf{r})$ and each possible ordering of all other field distributions with some non-zero spatial overlap. As such, including even a modest number of phantom channels can dramatically increase the computation resources needed to propagate the field numerically.

And so instead we seek a local basis in which the fields within the nonlinear region -- here taken to be the ring and the waveguide region coupled to the ring, where field enhancements will be large -- need only be described by a few modes at a given $\textbf{r}$. Luckily, such a mode basis can be constructed straightforwardly by considering linear combinations of the asymptotic-in(out) modes. Indeed, since the coupling between the loss channels and the ring and waveguide are equivalent to a beam-splitter-like interaction, rather than defining our mode basis in terms of steady-state classical solutions to the system with inputs (outputs) in a single input (output) channel as was done for the asymptotic-in(out) basis, we can consider modes defined with inputs in multiple channels, tuned such that destructive interference occurs almost everywhere within the nonlinear region. 

\begin{figure*}
    \centering
    \includegraphics[width=\textwidth]{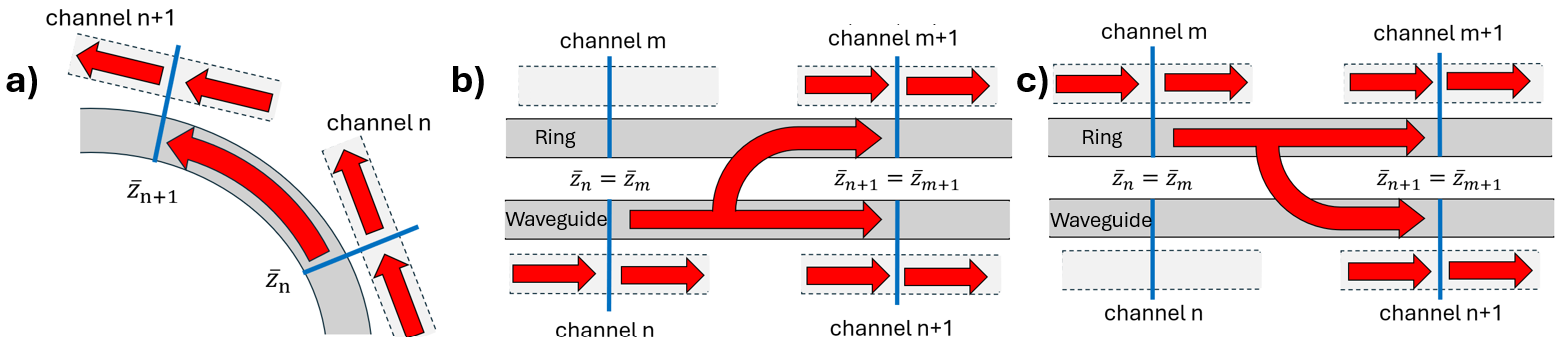}
    \caption{(a) Diagram of the energy flow for a local basis ring mode in which inputs from two adjacent phantom channels are tuned to restrict the non-zero field support within the ring to the area between the respective coupling points. (b) Sample local basis mode pair in the ring and waveguide coupling region with an initial input from a phantom channel coupled to the waveguide. (c) Sample local basis mode pair in the ring and waveguide coupling region with an initial input from a phantom channel coupled to the ring coupling region.}
    \label{fig:LocalBasisDiagrams}
\end{figure*}

As an example, consider the scenario depicted in Fig. \ref{fig:LocalBasisDiagrams}a, in which some classical CW field is incident from the $n^{th}$ and $(n+1)^{th}$ loss channel inputs (both of which are coupled to the ring outside the coupling region with the waveguide). By tuning the amplitude and phase of the input in the $(n+1)^{th}$ channel relative to the amplitude and phase of the input in the $n^{th}$ channel, one can achieve complete destructive interference within the ring at the point immediately after the coupling point for the $(n+1)^{th}$ channel. As a result, the only region of the ring that would have a non-zero field support would be the region between the $n^{th}$ and $(n+1)^{th}$ coupling point. However, the field distribution for the mode constructed through such a steady state solution to equation (\ref{eq:EOMofLinearFields}) can be interpreted as a linear combination of $\textbf{D}?{asy-in(out)}_{J,k,n}(\textbf{r})$ and $\textbf{D}?{asy-in(out)}_{J,k,n+1}(\textbf{r})$. Indeed, one can show that
\begin{widetext}
\begin{equation} \label{eq:SampleLocalBasisDisplacementField}
\begin{split}
    & \textbf{D}?{asy-in}_{J,k,n}(\textbf{r}) - \frac{\kappa?{ph}_{J,n}}{\kappa?{ph}_{J,n+1}} \sigma?{ph}_{J, n+1} \textbf{D}?{asy-in}_{J,k,n+1}(\textbf{r}) e^{i\Delta k_J \left( l_J(\bar{z}_{n+1}) - l_J(\bar{z}_{n}) \right)} e^{i k_J' (\bar{z}_{n+1} - \bar{z}_n)} \\
    & \qquad = \begin{cases}
        \textbf{d}_{J,n}?{ph}(x,y)e^{i k_{[n]} z} & \textit{in the } n^{th} \textit{ loss channel with } z < 0 \\
        -\frac{\kappa?{ph}_{J,n}}{\kappa?{ph}_{J,n+1}} \sigma?{ph}_{J, n+1}\textbf{d}_{J,n+1}?{ph}(x,y) e^{i\Delta k \left( l_J(\bar{z}_{n+1}) - l_J(\bar{z}_{n}) \right)} \\
        \qquad \qquad \qquad \times e^{i k' (\bar{z}_{n+1} - \bar{z}_n)} e^{i k_{[n+1]} z} & \textit{in the } (n+1)^{th} \textit{ loss channel with } z < 0 \\
        i\kappa?{ph}_{J,n} \textbf{d}?{asy}_{J,1}(x,y) e^{i\Delta k_J \left(l_J(z) - l_J(\bar{z}_n) \right)} e^{i k_J'(z - \bar{z}_n)} & \textit{in the ring with } \bar{z}_n < z < \bar{z}_{n+1} \\
        \sigma?{ph}_{J,n}\textbf{d}_{J,n}?{ph}(x,y)e^{i k_{[n]} z} & \textit{in the } n^{th} \textit{ loss channel with } z > 0 \\
        -\frac{\kappa?{ph}_{J,n}}{\kappa?{ph}_{J,n+1}} \textbf{d}_{J,n+1}?{ph}(x,y)  e^{i\Delta k \left( l_J(\bar{z}_{n+1}) - l_J(\bar{z}_{n}) \right)} \\
        \qquad \qquad \qquad \times e^{i k' (\bar{z}_{n+1} - \bar{z}_n)} e^{i k_{[n+1]} z} & \textit{in the } (n+1)^{th} \textit{ loss channel with } z > 0 \\
        0 & \textit{otherwise}
    \end{cases}
\end{split}
\end{equation}
\end{widetext}
where $\textbf{d}?{ph}_{J,n}(x,y)$ gives the field distribution dependence in the direction perpendicular to the field propagation in the $n^{th}$ phantom channel, with $k_{[n]}$ being the wavevector in this channel corresponding to $\omega = \omega_J + v_J \Delta k_J$.

Generalizing the process above, one can construct a local basis for the displacement field,
\begin{equation} \label{eq:LocalBasisDiplacementField}
\begin{split}
    \textbf{D}(\textbf{r}, t) &= \sum_{J,n} \int_{R(J)} dk \sqrt{\frac{\hbar \omega_J}{4\pi}} \textbf{D}?{loc}_{J,k, n}(\textbf{r}) \\
    & \qquad \qquad \qquad \qquad \times a?{loc}_{J,n}(k,t) + H.c.,
\end{split}
\end{equation}
in which each of the $\textbf{D}?{loc}_{J,k, n}(\textbf{r})$ are constructed by starting with $\textbf{D}?{asy-in}_{J,k, n}(\textbf{r})$, then adding the field distribution(s) corresponding to the next adjacent input(s), tuned to restrict the non-zero field support within the nonlinear region to the area between the respective inputs. 

Such a construction within the coupling region is more complicated than in the area outside the coupling region, and for an arbitrary placement of loss channels the field support may extend beyond the nearest neighbor phantom channel coupling point due to the fields oscillating between the ring and the waveguide. However, upon fixing the phantom channel coupling points within the ring, one can choose to place a phantom channel in the waveguide coupling region at the same $z$ position of each of the phantom channels in the ring coupling region. In this way, for $\bar{z}_n$ and $\bar{z}_{n+1}$ in the coupling region, one can construct a pair of modes, each of which having a non-zero support in the nonlinear region between the points $\bar{z}_n$ and $\bar{z}_{n+1}$, but contained within the ring and waveguide. Such a scheme is shown in Fig. \ref{fig:LocalBasisDiagrams}b and \ref{fig:LocalBasisDiagrams}c. We emphasize here that within the ring the locations of the phantom channel coupling points are still taken to be arbitrary, allowing for any number of coupled channels with arbitrary spacing between their coupling points.

This alignment of the waveguide phantom channels is convenient as, in the limit when the number of phantom channels is taken to be very large, each local basis mode effectively describes the field at a given $z$ point within the ring or waveguide coupling region. For completeness, in Appendix \ref{sec:appendix2} we include sample field distributions for the modes confined to the coupling region.  

Importantly for such a local basis, within the nonlinear region the field distribution $\textbf{D}?{loc}_{J,k, n}(\textbf{r})$ is either disjoint with all other $\textbf{D}?{loc}_{J',k', n'}(\textbf{r})$ for $n \neq n'$ (in the case when $\textbf{D}?{loc}_{J,k, n}(\textbf{r})$ is zero at all points within the ring and waveguide coupling region), or has a non-zero overlap with $\textbf{D}?{loc}_{J',k', n'}(\textbf{r})$ for only one $n'$ with $n' \neq n$ (when $\textbf{D}?{loc}_{J,k, n}(\textbf{r})$ is non-zero at some point in the ring and waveguide coupling region). Indeed, the product $\textbf{D}?{loc}_{J,k, n}(\textbf{r}) \textbf{D}?{loc}_{J',k', n'}(\textbf{r})$ can be non-zero only for $n$ and $n'$ such that $\bar{z}_n = \bar{z}_{n'}$ when $\textbf{r}$ corresponds to a point within the nonlinear region.

Letting $\mathcal{L}?{in}_{J,k}$ be the matrix defining the transformation between the field distributions $\textbf{D}?{loc}_{J,k, n}(\textbf{r})$ and $\textbf{D}?{asy-in}_{J,k, n}(\textbf{r})$ such that 
\begin{equation}
    \begin{bmatrix}
    \textbf{D}?{loc}_{J,k, 0}(\textbf{r}) \\
    \textbf{D}?{loc}_{J,k, 1}(\textbf{r}) \\
    \vdots \\
    \textbf{D}?{loc}_{J,k, N_L}(\textbf{r})
    \end{bmatrix}
    = \mathcal{L}?{in}_{J,k}
    \begin{bmatrix}
    \textbf{D}?{asy-in}_{J,k, 0}(\textbf{r}) \\
    \textbf{D}?{asy-in}_{J,k, 1}(\textbf{r}) \\
    \vdots \\
    \textbf{D}?{asy-in}_{J,k, N_L}(\textbf{r})
    \end{bmatrix},
\end{equation} 
it follows that for equations (\ref{eq:AsymptoticDisplacementField}) and (\ref{eq:LocalBasisDiplacementField}) to be equivalent we must have 
\begin{equation} \label{eq:OperatorTransformation}
    \begin{bmatrix}
        a?{loc}_{J,0}(k,t) \\
        a?{loc}_{J,1}(k,t) \\
        \vdots \\
        a?{loc}_{J,N_L}(k,t)
    \end{bmatrix} 
    = \left( \mathcal{L}?{in-1}_{J,k} \right)^T
    \begin{bmatrix}
        a?{in}_{J,0}(k,t) \\
        a?{in}_{J,1}(k,t) \\
        \vdots \\
        a?{in}_{J,N_L}(k,t)
    \end{bmatrix}.
\end{equation}

In general, the transformation $\mathcal{L}?{in}_{J,k}$ is not unitary and thus, unlike the $a?{in}_{J,n}(k,t)$, the operators $a?{loc}_{J,n}(k,t)$ do not satisfy the standard bosonic commutations relations. However, using equation (\ref{eq:OperatorTransformation}), it follows that
\begin{equation} \label{eq:LocalBasisCommutators}
\begin{split}
    \left[ a?{loc}_{J,n}(k,t), a?{loc}_{J',n'}(k',t) \right] &= 0 = \left[ a^{\textnormal{loc} \dagger}_{J,n}(k,t), a^{\textnormal{loc} \dagger}_{J',n'}(k',t) \right] \\
    \left[ a?{loc}_{J,n}(k,t), a^{\textnormal{loc} \dagger}_{J',n'}(k',t) \right] &= \delta_{J, J'} \delta(k - k') \\
    & \times \sum_{m = 0}^{N_L} \left[ \mathcal{L}^{\textnormal{in} -1}_{J',k'} \right]^{*}_{n',m} \left[ \mathcal{L}^{\textnormal{in} -1}_{J,k} \right]_{n,m}
\end{split}
\end{equation}

Of course, since each asymptotic-out mode is uniquely determined by an output in a single channel, then one could have instead constructed the local basis field distributions, $\textbf{D}?{loc}_{J,k, n}(\textbf{r})$, in terms of the asymptotic-out field distributions, $\textbf{D}?{asy-out}_{J,k, n}(\textbf{r})$. In this way one could define the transformation $\mathcal{L}?{out}_{J,k}$ and equivalently solve for the local basis commutators in terms of $\mathcal{L}^{\textnormal{out} -1}_{J,k}$.

We conclude this section with a few final notes about the local basis introduced here. First, while the asymptotic-in(out) basis and the coefficients relating the field distributions in these bases to the local basis were derived under the approximation of the ring being equivalent to a straight waveguide of finite length, this assumption is not necessary. Rather, being able to construct a local basis in this way comes as a result of approximating the coupling between the loss channels and the ring or waveguide as occurring at a single point, with the loss channels stimulating the same single spatial mode of the coupled ring and waveguide system. As such, even when allowing for bending in the ring, we can let $\xi$ denote the location along the circumference of the ring and $\textbf{r}_{\perp}$ the coordinates in the plane perpendicular to the propagation direction, and expand the asymptotic-in(out) field distributions as
\begin{equation}
    \textbf{D}?{asy-in(out)}_{J,k,n}(\textbf{r}) = \textbf{d}?{in(out)}_{J,k}(\textbf{r}_{\perp}; \xi) h?{in(out)}_{J,k,n}(\xi)
\end{equation}
where $\textbf{d}?{in(out)}_{J}(\textbf{r}_{\perp}; \xi)$ is independent of $n$. So one needs only to know the specifics of $h?{in(out)}_{J,k,n}(\xi)$, which can be computed numerically, to determine the entries of $\mathcal{L}?{in(out)}_{J,k}$ used to define the local basis. 

Consequentially, we have the freedom to include effects arising from any particular structure, such as the variation of the local index of refraction, loss, and nonlinear strength due to the distortion of the field distribution in the bent regions relative to those in the straight section. We can also extend the current treatment to include higher order spatial modes, with the change in curvature in the ring resulting in coupling of the primary spatial mode to higher order modes in the resonator. Furthermore, back-scattering of the fields can be constructed by allowing for counter-propagating wave modes in the waveguide and phantom channels. Compared to a more standard coupled-mode approach, which could include a phenomenological treatment of such effects, the asymptotic fields approach allows us to treat these locally with the local basis expansion providing a convenient way to relate dynamics of the full system to that contained between adjacent pairs of loss channels.

Finally, we note that while the derivation above has focused on a single microring resonator side-coupled to a channel waveguide, the approach can be generalized to a variety of integrated photonic elements. Indeed, the distribution of phantom channel along the length of the interaction region comes naturally as a result of discretizing the scattering loss. Regardless of the structure geometry, one constructs the local basis modes by adjusting the strength of the point coupled phantom channels inputs to control the field interference and achieve non-zero field only between two consecutive coupling points (see Fig. \ref{fig:LocalBasisDiagrams}).

\section{Nonlinear Interaction} \label{sec:NonlinearInteraction}
Having set out the mode basis for the displacement fields, we now introduce the Hamiltonian to describe our system, and derive the equations of motion for each of the operators $a?{loc}_{J,n}(k,t)$. In particular, we write the full Hamiltonian as
\begin{equation}
    H = H_L + H_{NL},
\end{equation}
where $H_L$ denotes the linear contributions to the Hamiltonian, with $H_{NL}$ containing the nonlinearity. The equations of motion for each of the operators in the Heisenberg picture are then given by
\begin{equation} \label{eq:EOMdefinition}
    \frac{\partial}{\partial t} a?{loc}_{J, n}(k,t) = \frac{1}{i\hbar} \left[ a?{loc}_{J, n}(k,t), H_L + H_{NL} \right].
\end{equation}
However, from equations (\ref{eq:LinearEvolutionKoperator}) and (\ref{eq:OperatorTransformation}) it follows that the linear term of the equation of motion is simply
\begin{equation}
    \frac{1}{i\hbar} \left[ a?{loc}_{J, n}(k,t), H_L \right] = -i \omega_{J,k} a?{loc}_{J, n}(k,t).
\end{equation}
Hence, to generate the full coupled equations of motion for the system, we need only to specify the nonlinear interactions.

For our current treatment, we are interested in the generation of photon pairs through spontaneous four-wave mixing processes, and as such will include nonlinear terms up to third order in the displacement field. Furthermore, we consider the ring and waveguide made from material with negligible second-order nonlinearity such that we can write the nonlinear Hamiltonian as \cite{Quesada:22}
\begin{equation} \label{eq:GeneralNonlinearHamiltonian_preExpansion}
    H_{NL} = -\frac{1}{4\epsilon_0} \int d\textbf{r} \Gamma_{(3)}^{ijkl}(\textbf{r}) D^i(\textbf{r},t) D^j(\textbf{r},t) D^k(\textbf{r},t) D^l(\textbf{r},t) 
\end{equation}
where $\Gamma_{(3)}^{ijkl}(\textbf{r})$ is a third order nonlinear tensor, with the summation over $i, j, k, l \in \{\hat{x}, \hat{y}, \hat{z}\}$ implicit. The integral over $\textbf{r}$ is taken along the nonlinear region with the $z$ component ranging from $z=0$ to $z=L_r$ in the ring and $z=0$ to $z=L_c$ within the waveguide. 

Expanding the displacement field in terms of the local basis as in equation (\ref{eq:LocalBasisDiplacementField}), we can write
\begin{equation}
    \textbf{D}?{loc}_{J, k, n}(\textbf{r}) = \textbf{d}_{J, \tau}?{asy}(x,y) h?{loc}_{J,k,\tau}(z) e^{ik_J'z},
\end{equation}
and, assuming the resonances of interest lie close together, we include only terms of the form $a^{\dagger}_1 a^{\dagger}_2 a_3 a_4$ that can lead to an energy preserving interaction. Consequentially, writing $\vec{J} = \left( J_1, J_2, J_3, J_4 \right)$ and $\vec{n} = \left( n_1, n_2, n_3, n_4 \right)$, the nonlinear Hamiltonian becomes
\begin{equation} \label{eq:GeneralNonlinearHamiltonian}
\begin{split}
    H_{NL} &= -\frac{\hbar}{(2\pi)^2} \sum_{\vec{J}, \vec{n}} \int d\textbf{k} \int dz \Lambda_{\vec{J}}^{\tau} \mathcal{J}^{\tau}_{\vec{J}, \vec{k}, \vec{n}}(z) e^{-i \delta k^0_{\vec{J}}z} \\
    & \qquad \qquad \times a^{\textnormal{loc} \dagger}_{J_1, n_1}(k_1,t) a^{\textnormal{loc} \dagger}_{J_2, n_2}(k_2,t) a?{loc}_{J_3, n_3}(k_3,t) \\
    & \qquad \qquad \times a?{loc}_{J_4, n_4}(k_4,t),
\end{split}
\end{equation}
for $d \textbf{k} = dk_1 dk_2 dk_3 dk_4$ and $\mathcal{J}^{\tau}_{\vec{J}, \vec{k}, \vec{n}}(z)$ representing the overlap of the slowly varying field envelopes, defined as
\begin{equation}
\begin{split}
    \mathcal{J}^{\tau}_{\vec{J}, \vec{k}, \vec{n}}(z) &= h?{loc *}_{J_1, k_1, n_1}(z) h?{loc *}_{J_2, k_2, n_2}(z) h?{loc}_{J_3, k_3, n_3}(z)\\
    & \qquad \times h?{loc}_{J_4, k_4, n_4}(z).
\end{split}
\end{equation}
Additionally, we have introduced the nonlinear strength, $\Lambda_{\vec{J}}^{\tau}$, given by
\begin{equation}
    \Lambda_{\vec{J}}^{\tau} = \frac{1}{2} \hbar \omega_{\vec{J}} \gamma_{NL}^{\vec{J}, \tau} v_{\vec{J}}^2,
\end{equation}
where the nonlinear parameter $\gamma_{NL}^{\vec{J} , \tau}$ is related to the overlap over the tangential field distributions as
\begin{equation}
\begin{split}
   \gamma_{NL}^{\vec{J}, \tau} &= \frac{3 \omega_{\vec{J}}}{4 \epsilon_0 v_{\vec{J}}^2} \int \int dx dy \Gamma^{ijkl}_{(3), \tau}(x,y) \textbf{d}^{\textnormal{asy}, i*}_{J_1, \tau}(x,y) \\
   & \qquad \qquad \times \textbf{d}^{\textnormal{asy},j*}_{J_2, \tau}(x,y) \textbf{d}^{\textnormal{asy}, k}_{J_3, \tau}(x,y) \textbf{d}^{\textnormal{asy}, l}_{J_4, \tau}(x,y), 
\end{split}
\end{equation}
and $\omega_{\vec{J}}$ and $v_{\vec{J}}$ are each defined as
\begin{equation}
\begin{split}
    \omega_{\vec{J}} &= \left( \omega_{J_1} \omega_{J_2} \omega_{J_3} \omega_{J_4} \right)^{1/4} \\
    v_{\vec{J}} &= \left( v_{J_1} v_{J_2} v_{J_3} v_{J_4} \right)^{1/4}
\end{split}
\end{equation}
Note that $\Gamma^{ijkl}_{(3)}(\textbf{r}) = \Gamma^{ijkl}_{(3), \tau}(x, y)$ is a function of the standard nonlinear tensor $\chi^{ijkl}_{(3)}(\textbf{r})$ which, upon taking each of the fields to be polarized perpendicular to the ring and waveguide plane, we have assumed does not vary with the $z$ coordinate \cite{Quesada:22}. Finally, $\delta k^0_{\vec{J}} = k_{J_1}' + k_{J_2}' - k_{J_3}' - k_{J_4}'$ gives the detuning of the resonance centers.

Writing the commutators of the creation and annihilation operators in the local basis by
\begin{equation}
    \left[ a?{loc}_{J, n}(k, t), a^{\textnormal{loc} \dagger}_{J,n'}(k', t) \right] = \mathcal{C}^J_{n, n'}(k) \delta \left( k - k' \right),
\end{equation}
(recall equation (\ref{eq:LocalBasisCommutators})), we can then use the nonlinear Hamiltonian from (\ref{eq:GeneralNonlinearHamiltonian}) in equation (\ref{eq:EOMdefinition}) to find the equations of motion of the local basis operators, leading to
\begin{equation} \label{eq:GeneralEquationOfMotion}
\begin{split}
    \frac{\partial}{\partial t} a?{loc}_{J, n}(k, t) &= -i\omega_{J, k} a?{loc}_{J, n}(k, t) \\
    & \qquad + \frac{i}{(2\pi)^2} \sum_{\vec{J}, \vec{n}} g_{\vec{J}} \Lambda_{\vec{J}}^{\tau} \int d \textbf{k} \int dz \\
    & \qquad \qquad \times \mathcal{C}^J_{n, n_1}(k) \mathcal{J}^{\tau}_{\vec{J}, \vec{k}, \vec{n}} (z) a^{\textnormal{loc} \dagger}_{J_2, n_2}(k_2,t) \\
    & \qquad \qquad \times a?{loc}_{J_3, n_3}(k_3,t) a?{loc}_{J_4, n_4}(k_4,t) \\
    & \qquad \qquad \times e^{-i \delta k^0_{\vec{J}} z} \delta_{J, J_1} \delta (k - k_1).
\end{split}
\end{equation}
Here $g_{\vec{J}}$ is a combinatorial factor given by 
\begin{equation}
    g_{\vec{J}} = \begin{cases}
        1 & \textit{when } J_1 \neq J_2 \\
        2 & \textit{when } J_1 = J_2
    \end{cases},
\end{equation}
which takes into account the ordering of the $J_1$ and $J_2$ terms in the Hamiltonian.

The equations of motion presented in equation (\ref{eq:GeneralEquationOfMotion}) are in the most general form for a lossy ring coupled to a waveguide, with an arbitrary number of phantom channels and an arbitrary spacing between them. However, in many situations of practical interest we can develop the equations of motion into a form more convenient for numerical solutions. To investigate this, we consider first the overlap function $\mathcal{J}^{\tau}_{\vec{J}, \vec{k}, \vec{n}}(z)$ for $n_1 = n$, where $n$ corresponds to a field distribution confined to the ring outside the coupling region as in equation (\ref{eq:SampleLocalBasisDisplacementField}). For this choice of $n$, from equation (\ref{eq:SampleLocalBasisDisplacementField}) it follows that the only $k$ dependence in $h?{loc}_{J, k, n}(z)$ comes from the $z$ dependent phase factor. Hence, we can write
\begin{equation}
    \begin{split}
        \mathcal{J}^{\tau}_{\vec{J}, \vec{k}, \vec{n}}(z) &= \delta_{n, n_1} \delta_{n, n_2} \delta_{n, n_3} \delta_{n, n_4} \\
        & \quad \times \tilde{h}?{loc *}_{J_1, n} \tilde{h}?{loc *}_{J_2, n} \tilde{h}?{loc}_{J_3, n} \tilde{h}?{loc}_{J_4, n} e^{-i \delta k_{\vec{J}} (z-\bar{z}_n)},
    \end{split}
\end{equation}
where $\tilde{h}?{loc}_{\vec{J}, \vec{n}}$ is a $k$-independent amplitude and the phase is given by
\begin{equation}
\begin{split}
    \delta k_{\vec{J}} &= \frac{v_{J_1}}{u_{J_1}}(k_1 - k_{J_1}) + \frac{v_{J_2}}{u_{J_2}}(k_2 - k_{J_2}) \\
    & \quad - \frac{v_{J_3}}{u_{J_3}}(k_3 - k_{J_3}) - \frac{v_{J_4}}{u_{J_4}}(k_4 - k_{J_4}).
\end{split}
\end{equation}
By choosing to equally distribute the phantom channels around the ring such that $\bar{z}_{n+1} - \bar{z}_n = L_r/N_L^R$, with $N_L^R$ the number of phantom channels coupled to the ring, we can then write the integral over the $z$ dependent components of the equations of motion as
\begin{equation}
\begin{split}
    & \int_{\bar{z}_n}^{\bar{z}_{n+1}} dz \mathcal{J}^{\tau}_{\vec{J}, \vec{k}, \vec{n}}(z) e^{i\delta k^0_{\vec{J}} z}  \\
    & \qquad = \delta_{n, n_1} \delta_{n, n_2} \delta_{n, n_3} \delta_{n, n_4} \tilde{h}^{\textnormal{loc} *}_{J_1, n} \tilde{h}^{\textnormal{loc} *}_{J_2, n} \tilde{h}?{loc}_{J_3, n} \tilde{h}?{loc}_{J_4, n} \\
    & \qquad \qquad \times \frac{e^{i(\delta k^0_{\vec{J}} + \delta k_{\vec{J}}) \frac{L_r}{N_L^R}} - 1}{\delta k^0_{\vec{J}} + \delta k_{\vec{J}}} e^{i\delta k^0_{\vec{J}} \bar{z}_n}.
\end{split}
\end{equation}
Here the $\vec{k}$ dependent factor takes into account the phase mismatch between the fields along the interval between the $\bar{z}_n$ and $\bar{z}_{n+1}$. In the case of a large ring made from the same material as the waveguide, one would expect that $v_J/u_J \cong 1$ for each resonance, with pair generation being appreciable within a few linewidths from the center of the signal and idler resonances. Denoting by $\Gamma_J$ the half-width at half maximum of the resonance $J$ and $\delta \nu$ the free spectral range (FSR), then for the purpose of squeezed light generation we will typically be interested in resonators with a high enough finesse that $\Gamma_J \ll \delta \nu$. Hence, even for a modest choice of $N_L$ we have $\delta k_{\vec{J}} \frac{L_r}{N_L^R} \ll 1$. On the other hand, the factor $\delta k^0_{\vec{J}}$ is related to the shifts in the resonance peaks due to group velocity dispersion. Of course, one would want to keep this small for desirable nonlinear interactions, but even when it is large, one can choose $N_L$ such that $\delta k^0_{\vec{J}}\frac{L_r}{N_L^R} \ll 1$. In this case, we can make the approximation
\begin{equation} \label{eq:zIntegralSimplification}
    \int_{\bar{z}_n}^{\bar{z}_{n+1}} dz \mathcal{J}^{\tau}_{\vec{J}, \vec{k}, \vec{n}}(z) e^{i\delta k^0_{\vec{J}} z}  \cong \tilde{\mathcal{J}}^{\tau}_{\vec{J}, \vec{n}} e^{i \delta k^0_{\vec{J}}\bar{z}_n},
\end{equation}
for some $\tilde{\mathcal{J}}^{\tau}_{\vec{J}, \vec{n}}$ which is independent of $\vec{k}$.

Within this approximation, along the length between adjacent loss channels all terms included in the $\vec{J}$ sum are perfectly phase matched for all values of $k$ in the given resonance ranges. The global phase mismatch of the fields along the whole length of the ring is still taken into account; but due to the form of the local basis it is now folded into the basis transformations, and consequentially its effects are captured by the commutators.

The overlaps between the field distributions that have some non-zero support within the coupling region can be more complicated, since in addition to the $k$ dependent phase factor there is also a $k$ dependence in the envelope frequency $\alpha_k$, as well as in $\gamma_J(k)$ and $\bar{\sigma}(k)$; both of these depend on $k$ through their dependence on $\alpha_{k}$. However, if the difference in the group velocity in the waveguide and the ring coupling region is small, in particular when $\left| 1 - \frac{v_J}{u_J} \right| \ll 1$, the effect of the detuning from the central resonance on $\alpha_k$ tends to be small. Hence we can make the approximations
\begin{equation}
\begin{split}
    \alpha_k &\rightarrow \alpha_J \equiv \alpha_{k_J}, \\
    \gamma_J(k) &\rightarrow \gamma_J \equiv \gamma_J(k_J), \\
    \bar{\sigma}(k) &\rightarrow \bar{\sigma}_J \equiv \bar{\sigma}_J(k_J),
\end{split} 
\end{equation}
after which the only $k$ dependence in $\mathcal{J}^{\tau}_{\vec{J}, \vec{k}, \vec{n}}(z)$ comes from the phase factor. Then we can simplify the $z$ integral for any choice of $\vec{J}$, $\vec{k}$, $\vec{n}$, and $\tau$ in the same way as (\ref{eq:zIntegralSimplification}), and write the equations of motion as
\begin{equation} \label{eq:SimplifiedEOM}
    \begin{split}
    \frac{\partial}{\partial t} a?{loc}_{J, m}(k, t) &= -i\omega_{J, k} a^{loc}_{J, m}(k, t) \\
    & \quad + \frac{i}{(2\pi)^2} \sum_{\vec{J}, \vec{n}, \tau} \tilde{\Lambda}_{\vec{J}}^{\tau, \vec{n}, m} (k) \int dk_2 dk_3 dk_4 \\
    & \quad \qquad \times a^{\textnormal{loc} \dagger}_{J_2, n_2}(k_2,t) a?{loc}_{J_3, n_3}(k_3,t) \\
    & \quad \qquad \times a?{loc}_{J_4, n_4}(k_4,t) e^{-i \delta k^0_{\vec{J}} \bar{z}_n},
\end{split}
\end{equation}
with $\tilde{\Lambda}^{\tau, \vec{n}, m}_{\vec{J}}(k) = g_{\vec{J}}\Lambda^{\tau}_{\vec{J}} \tilde{\mathcal{J}}^{\tau}_{\vec{J}, \vec{n}}\mathcal{C}^{J_1}_{m, n_1}(k)$ denoting the effective nonlinear mixing of the resonances in $\vec{J}$, modes $\vec{n}$, within the region $\tau$ (waveguide or ring).

\begin{figure}
    \centering
    \includegraphics[width=0.45\textwidth]{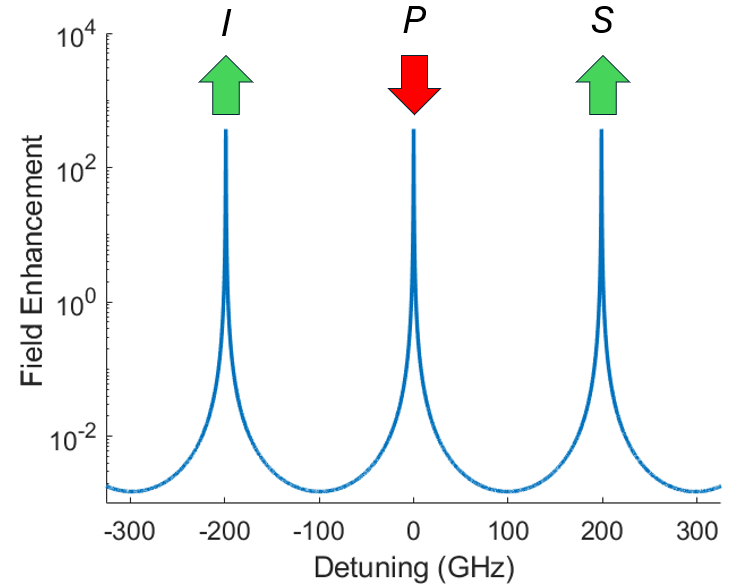}
    \caption{Spectrum of the three resonance system considered in the single pump, non-degenerate SFWM example for a ring of length $L_r = 2\pi \times 120$ $\mu$m and a finesse of $\mathcal{F}=780$. Resonances are labelled I, P, and S corresponding to the idler, pump, and signal respectively.}
    \label{fig:ThreeResonances}
\end{figure}

Note that while the approximation introduced in equation (\ref{eq:zIntegralSimplification}) leads to a simplification of the operator dynamics in equation (\ref{eq:SimplifiedEOM}), this also places a limit on the minimum number of loss channels required. Furthermore, such a limit depends on the finesse of the resonator in question. A low finesse results in appreciable pair generation within a broader fraction of the FSR, thus necessitating more loss channels to ensure the phase mismatch remains small along the regions between coupling points. For the case of a large resonator of length $L_r = 2\pi \times 120$ $\mu$m and finesse ranging from $\mathcal{F} = 10^1 - 10^3$, we find that $N_L^R = 10-20$ is more than sufficient for convergence, with additional loss channels distributing the energy radiated to the loss channels over more modes but having a negligible effect on the evolution of the output fields.

\section{Field Propagation: Single Pump SFWM} \label{sec:SampleCalculation}
To demonstrate the use of the equations of motion (\ref{eq:SimplifiedEOM}) to numerically solve for the time evolution of the relevant fields, we consider the simple example of single pump, non-degenerate SFWM, with S, P, and I denoting the signal, pump, and idler resonances respectively (see Fig. \ref{fig:ThreeResonances}). We take the pump resonance to be driven by a classical pump, with pairs of photons being generated in the signal and idler resonances, but still sufficiently weak that pump depletion, as well as self-phase modulation (SPM) of, and cross-phase modulation (XPM) from, the signal and idler fields can be safely neglected. Of course, in general there will be other signal and idler fields generated in resonances further from the pump, but we consider the pump sufficiently weak that any parasitic processes on our signal and idler fields of interest involving the mixing of those fields with the pump field can be neglected. Then the nonlinear interactions of interest will be: single pump SFWM, SPM of the pump field, and XPM of the signal and idler fields by the pump field.

Since SPM is the only interaction affecting the pump we begin by fully solving for the time evolution of the pump field, and then use this to seed the generation of pairs in the signal and idler resonance. To do this, we first restrict the sum over $\vec{J}$ to only $\vec{J} = (P,P,P,P)$, then make the replacement 
\begin{equation}
    a?{loc}_{P, n}(k, t) \rightarrow \alpha?{loc}_{P, n}(k,t) \equiv \langle a?{loc}_{P, n}(k, t) \rangle  
\end{equation}
in the equations of motion for the pump, leading to
\begin{equation} \label{eq:PumpEOM}
    \begin{split}
    \frac{\partial}{\partial t} \alpha?{loc}_{P, m}(k, t) &= -i\omega_{J, k} \alpha?{loc}_{P, m}(k, t) \\
    & \quad + \frac{i}{(2\pi)^2} \sum_{\vec{n}, \tau} \tilde{\Lambda}_{PPPP}^{\tau, \vec{n}, m}(k) \int dk_2 dk_3 dk_4 \\
    & \qquad \times \alpha^{\textnormal{loc} *}_{P, n_2}(k_2,t) \alpha?{loc}_{P, n_3}(k_3,t) \alpha?{loc}_{P, n_4}(k_4,t).
\end{split}
\end{equation}
Unlike equations (\ref{eq:SimplifiedEOM}), this is now a system of nonlinear differential equations of classical variables that can readily be solved using a number of approaches, such as linear multi-step methods \cite{linearMultiStep}. The initial conditions for the solution of each of the $\alpha_{P,n}?{loc}(k,t)$ will in general consist of a pulse incident from the waveguide, and vacuum in all other input channels. We choose an initial time $t_i$ such that the pulse has yet to reach the ring, define the pump field in the asymptotic-in basis, then use equation (\ref{eq:OperatorTransformation}) to convert this into the local basis, $\alpha?{loc}_{P,n}(k,t)$. Using equation (\ref{eq:PumpEOM}), we can then propagate the fields forward in time to solve for $\alpha?{loc}_{P, n}(k,t)$ for all $t>t_i$.

With the solution for the pump field 
in hand, we turn to the generation and propagation of the signal and idler fields. Taking into account XPM from the pump and non-degenerate SFWM, from equations (\ref{eq:SimplifiedEOM}) the coupled differential equations become
\begin{widetext}
    \begin{equation}
        \begin{split}
            \frac{\partial}{\partial t} a?{loc}_{S, m}(k,t) &= -i\omega_{S, k}a?{loc}_{S, m}(k,t) + \frac{i}{(2\pi)^2} \sum_{\vec{n}, \tau} \int dk_2 dk_3 dk_4 \left\{ \tilde{\Lambda}^{\tau, \vec{n},m}_{SPSP}(k) \alpha^{\textnormal{loc} *}_{P, n_2}(k_2,t)\alpha?{loc}_{P, n_4}(k_4,t) a?{loc}_{S, n_3}(k_3,t) \right. \\
            & \qquad \qquad \qquad \qquad \qquad \qquad \qquad \qquad \qquad + \left. \tilde{\Lambda}^{\tau, \vec{n},m}_{SIPP}(k) \alpha^{\textnormal{loc} *}_{P, n_3}(k_3,t)\alpha^{\textnormal{loc} *}_{P, n_4}(k_4,t) a^{\textnormal{loc} \dagger}_{I, n_2}(k_2,t) e^{-i\delta \bar{k} \bar{z}_{n_1} } \right\} \\
            \frac{\partial}{\partial t} a?{loc}_{I, m}(k,t) &= -i\omega_{I, k}a?{loc}_{I, m}(k,t) + \frac{i}{(2\pi)^2} \sum_{\vec{n}, \tau} \int dk_2 dk_3 dk_4 \left\{ \tilde{\Lambda}^{\tau, \vec{n},m}_{IPIP}(k) \alpha^{\textnormal{loc} *}_{P, n_2}(k_2,t)\alpha?{loc}_{P, n_4}(k_4,t) a?{loc}_{I, n_3}(k_3,t) \right. \\
            & \qquad \qquad \qquad \qquad \qquad \qquad \qquad \qquad \qquad + \left. \tilde{\Lambda}^{\tau, \vec{n},m}_{SIPP}(k) \alpha?{loc *}_{P, n_3}(k_3,t)\alpha^{\textnormal{loc} *}_{P, n_4}(k_4,t) a^{\textnormal{loc} \dagger}_{S, n_2}(k_2,t) e^{-i\delta \bar{k} \bar{z}_{n_1} } \right\},
        \end{split}
    \end{equation}
\end{widetext}
where we have written
\begin{equation}
    \delta \bar{k} = \delta k^0_{SIPP} = k_S' + k_I' - 2k_P'.
\end{equation}

These equations can be written in a more compact form by first gathering all of the spatial mode operators as 
\begin{equation}
    \vec{a}?{loc}_{J}(k, t) = \left[ a?{loc}_{J, 0}(k,t), \dots, a?{loc}_{J, N_L}(k,t) \right]^T.
\end{equation}
Then we discretize the $k$ range for the signal and idler resonances into $N_k$ bins, with central wavevectors given by $\{ k^S_i \}_{i=1}^{N_k}$ and $\{ k^I_i \}_{i=1}^{N_k}$ for the signal and idler respectively. We can group these resonances together as
\begin{equation}
    \textbf{a}_i?{loc}(t) = 
    \begin{pmatrix} \vec{a}?{loc}_{S}(k_i^S, t) \\ \vec{a}?{loc}_{I}(k_i^I, t) \\ \vec{a}^{\textnormal{loc} \dagger}_{S}(k_i^S, t) \\ \vec{a}^{\textnormal{loc} \dagger}_{I}(k_i^I, t) 
    \end{pmatrix},
\end{equation}
and write the equations of motion for the coupled system as
\begin{equation} \label{eq:CompactEOM}
    \frac{\partial}{\partial t} \textbf{a}_i?{loc}(t) = i\mathcal{A}?L_i \textbf{a}_i?{loc}(t) + i\mathcal{A}?{NL}_i(t) \sum_{j=1}^{N_k} \textbf{a}_j?{loc}(t).
\end{equation}
Here the matrices $\mathcal{A}?L_i$ contain the time independent linear terms in the equations of motion, and the matrices $\mathcal{A}?{NL}_i(t)$ contain the time dependent nonlinear terms. Importantly, the nonlinear matrices $\mathcal{A}?{NL}_i(t)$ have no $j$ dependence as a result of equation (\ref{eq:zIntegralSimplification}).

From the form of the equations of motion in equation (\ref{eq:CompactEOM}), we can utilize a split step method to write the short time propagation of the signal and idler fields as the successive application of a linear propagation matrix (constructed from the $\mathcal{A}_i?{L}$) followed by the application of a nonlinear propagation matrix (formed from the $\mathcal{A}?{NL}_i(t)$). The linear evolution in this system is simple and just results in the addition of a $k$ dependent phase for each of the operators. In particular, for a time step $\Delta t$, the linear evolution is given by
\begin{equation} \label{eq:LinearEvo}
    a?{loc}_{J, m}(k,t) \rightarrow a?{loc}_{J, m}(k,t)e^{-i \omega_{J, k} \Delta t}.
\end{equation}
Or, grouping the operators as we have done in equation (\ref{eq:CompactEOM}), we can write the linear evolution of the fields as
\begin{equation}
    \textbf{a}?{loc}_i(t) \rightarrow U_i?L(\Delta t) \textbf{a}?{loc}_i(t),
\end{equation}
with each of the $U?L_i(\Delta t)$ being diagonal with phases given by equation (\ref{eq:LinearEvo}).

For the nonlinear step, the relevant coupled system of equations is given by
\begin{equation}
    \frac{\partial}{\partial t} \textbf{a}_i?{loc}(t) = i\mathcal{A}?{NL}_i(t) \sum_{j=1}^{N_k} \textbf{a}_j?{loc}(t).
\end{equation}
It can be shown (see Appendix \ref{sec:appendix3}) that, upon taking $\mathcal{A}?{NL}_i(t)$ to be constant over a short time interval $\Delta t$, the nonlinear propagation can be written as
\begin{equation} \label{eq:FullEvoSolution}
    \textbf{a}_i?{loc}(t) \rightarrow \textbf{a}_i?{loc}(t) + U?{NL}_{i}(t, t+\Delta t) \left[ \sum_j \textbf{a}_j?{loc}(t) \right],
\end{equation}
where, for each $i$, the matrices $U?{NL}_i(t, t + \Delta t)$ are given by
\begin{equation} \label{eq:NLevoMats}
    U?{NL}_i(t, t + \Delta t) = iA?{NL}_i(t) \sum_n \frac{\Delta t^{n+1}}{(n+1)!} \left( i \tilde{\mathcal{A}}?{NL}(t) \right)^n,
\end{equation}
for
\begin{equation}
    \tilde{\mathcal{A}}?{NL}(t) = \sum_i A?{NL}_i(t).
\end{equation}
In the case when the matrix $\tilde{\mathcal{A}}?{NL}(t)$ is invertible, this can simply be written as
\begin{equation}
    U?{NL}_i(t, t + \Delta t) = \mathcal{A}?{NL}_i(t) \left(\tilde{\mathcal{A}}?{NL}(t) \right)^{-1} \left[e^{i\Delta t \tilde{\mathcal{A}}?{NL}(t)} - \mathcal{I} \right]
\end{equation}

To construct the full evolution of the fields over a time step $\Delta t$, we allow the operators to evolve linearly over a time $\Delta t/2$, followed by a nonlinear evolution over a time $\Delta t$, and another linear evolution for a time $\Delta t/2$. With this, we can write the full evolution of the operators as 
\begin{widetext}
\begin{equation} \label{eq:FullTimeEvo}
    \textbf{a}_i?{loc}(t + \Delta t) = U?L_i(\Delta t) \textbf{a}_i?{loc}(t) + U?L_i(\Delta t/2) U?{NL}_i(t, t+\Delta t) \left[ \sum_j U?L_j(\Delta t/2) \textbf{a}_j?{loc}(t) \right]
\end{equation}
\end{widetext}

From this we can build up the time evolution from some initial time $t_i$ to an arbitrary final time $t_f = t_i + n_s\Delta t$ by successive application of $n_s$ short time propagations given by equation (\ref{eq:FullTimeEvo}). Specifically, letting $\mathcal{U}?{loc}(t, t+\Delta t)$ be the short time step matrix such that
\begin{equation}
    \begin{pmatrix}
    \textbf{a}?{loc}_1(t+\Delta t) \\
    \vdots \\
    \textbf{a}?{loc}_{N_k}(t+\Delta t)
    \end{pmatrix} = \mathcal{U}?{loc}(t, t+\Delta t) \begin{pmatrix}
    \textbf{a}?{loc}_1(t) \\
    \vdots \\
    \textbf{a}?{loc}_{N_k}(t)
    \end{pmatrix},
\end{equation}
we construct the full time evolution matrix as
\begin{equation}
    \mathcal{U}?{loc}(t_i, t_f) = \mathcal{U}?{loc}(t_f - \Delta t, t_f) \dots \mathcal{U}?{loc}(t_i, t_i + \Delta t)
\end{equation}

With all of the operators in the local basis in hand at the final time $t_f$, we can then transform into the asymptotic-out basis using the $\mathcal{L}?{out}_{J,k}$ matrices to solve for the full time evolution matrix $\mathcal{U}?{out}(t_i, t_f)$ satisfying
\begin{equation} \label{eq:FinalTimeAsyOut}
    \begin{pmatrix}
    \textbf{a}?{out}_1(t_f) \\
    \vdots \\
    \textbf{a}?{out}_{N_k}(t_f)
    \end{pmatrix} = \mathcal{U}?{out}(t_i, t_f) \begin{pmatrix}
    \textbf{a}?{out}_1(t_i) \\
    \vdots \\
    \textbf{a}?{out}_{N_k}(t_i)
    \end{pmatrix},
\end{equation}
which gives a convenient way to analyze the field output from each channel.

To conclude this section we discuss the form of the solution in equation (\ref{eq:FullTimeEvo}). Note that instead of the approach taken here we could have remained in the asymptotic-in(out) basis, and upon discretizing the $k$ ranges we would have gathered all the operators for each $J$, $n$, and $k$ into a vector $\textbf{b}?{in(out)}(t)$ such that the equations of motion could be written as
\begin{equation}
    \frac{\partial}{\partial t} \begin{pmatrix} \textbf{b}?{in(out)}(t) \\ \textbf{b}^{\textnormal{in(out)} \dagger}(t) \end{pmatrix} = i\mathcal{B}?{in(out)}(t) \begin{pmatrix} \textbf{b}?{in(out)}(t) \\ \textbf{b}^{\textnormal{in(out)} \dagger}(t) \end{pmatrix}.
\end{equation}
To solve these equations, performing a Suzuki-Trotter decomposition of the full time evolution matrix into the product of a number of short time steps would then have required the computation of the exponential $e^{i \textbf{B}?{in(out)}(t) \Delta t}$ for each time step. The difficulty with this approach is that the matrix $\mathcal{B}?{in(out)}(t)$ has size $2N_RN_k(N_L+1) \times 2N_RN_k(N_L+1)$, where $N_R$ is the number of non-classical resonances ($N_R=2$ in this example); even for modest values of $N_k$ and $N_L$, this can be very large. As such, the computational resources and time needed to propagate the fields directly in the asymptotic-in(out) basis can quickly become unmanageable. 

On the other hand, for the method presented in this section that employs the local basis, solving for the nonlinear evolution over a short time step involves an exponential of $\tilde{\mathcal{A}}?{NL}(t)$, which only has size $2N_R(N_L+1) \times 2N_R(N_L+1)$. The linear evolution is comparatively fast, and so there is a dramatic decrease in the run time compared to that required by a straightforward application of a Suzuki-Trotter decomposition in the asymptotic-in(out) basis.

More generally, the asymptotic scattering method developed here relies on calculating the short time propagation matrices, $\mathcal{U}?{loc}(t, t+\Delta t)$, at each discretized time step. Consequentially, it scales quadratically with the number of resonances, frequency-bin discretization, and number of phantom channels. Compared to a more standard coupled mode treatment \cite{Quesada:22, Seifoory, Helt:10, Vernon-PhysRevA.92.033840, Cui-PhysRevResearch.3.013199} which involves the computation of Green function matrices along the full interaction time, and as such also scales quadratically with the number of resonances and frequency resolution, this asymptotic method leads to longer computation times owing to the larger mode dimensionality introduced by the phantom channels, but enables a local treatment of the field evolution within the ring structure. Perturbative asymptotic field based methods \cite{Liscidini, Banic-PhysRevA.106.043707}, on the other hand, provide a semi-analytic treatment of the pair generation in the structure, but do not allow for the complete reconstruction of the generated field output needed for the squeezing and coherence function calculations that will be presented in the following section.

\section{Sample calculations}
\label{sec:Results}
In this section we present sample calculations of the single pump SFWM scenario described above. We consider a ring and waveguide made from the same material with the same cross section dimensions such that, for simplicity, we can take $k_J' = k_J$ and $u_J = v_J$ for all resonances $J$. Furthermore, we will take $R_{e} = 120$ $\mu$m to define the effective radius of the ring resonator such that $L_r = 2\pi R_{e}$, and consider a coupling region which extends over a length of $L_c = L_r/4$ with separation large enough that the ring is weakly coupled to the resonator ($\alpha_J L_c \ll \pi$). We will also fix the nonlinear parameter as $\gamma_{NL}^{\vec{J}, \tau} = 1.0$ $(mW)^{-1}$, which is in line with previous calculations of squeezing in silicon nitride microrings \cite{Vernon-PhysRevA.92.033840, Seifoory}. 

Additionally, we take the loss in the ring coupling region and the waveguide coupling region to be the same. Then in the limit when the number of phantom channels approaches infinity ($N_L \rightarrow \infty$), the field amplitude for the asymptotic-in mode corresponding to the waveguide input would be constructed in the same way as equation (\ref{eq:sampleClassicalSolution}) but with the substitution
\begin{equation}
    e^{\Delta k_J l_J(z)} \rightarrow e^{\Delta k_J l_J(z)} e^{-\zeta(z)/2}.
\end{equation}
Here $\zeta(z)$ is a monotonically increasing function of $z$ describing the amplitude attenuation along the system. As such, for $\textbf{r}$ in the ring outside the coupling region ($L_c < z \leq L_r$) and with the assumptions above, it follows that the slowly varying field amplitude for the waveguide input asymptotic-in mode is 
\begin{equation}
\begin{split}
    h?{in}_{J,k,1}(z) &= \frac{i \bar{\kappa}_J}{1 - \bar{\sigma}_J e^{-\zeta(L_r)/2} e^{\Delta k_J L_r}} e^{-\zeta(z)/2} e^{i\Delta k_J z} \\
    &= F_J(\Delta k_J) e^{-\zeta(z)/2} e^{i\Delta k_J z},
\end{split}
\end{equation}
where the effective intensity enhancement of the resonance $J$ near $k_J$ is given by 
\begin{equation}
    |F_J(\Delta k_J)|^2 = \frac{\bar{\kappa}_J^2}{1 + \bar{\sigma}_J^2 \xi^2 - 2\bar{\sigma}_J \xi \cos (\Delta k_J L_r)}
\end{equation}
for $\xi = e^{-\zeta(L_r)/2}$. From this, when the width of the resonance $J$ is small compared to the spacing between resonances $\delta \nu$, the half-width at half max of the resonance is approximately 
\begin{equation}
    \Gamma_J \cong \frac{v_J}{2 \pi} \frac{1 - \bar{\sigma}_J \xi}{\sqrt{\bar{\sigma}_J \xi}} \frac{1}{L}.
\end{equation}
To characterize the loss we introduce an escape efficiency, $\eta_{\textnormal{esc}}$, representing the fraction of power exiting the ring through the output waveguide. Within the limit (72), and summing over all the phantom waveguides, we find
\begin{equation}
    \eta_{\textnormal{esc}} \cong \frac{(1 - \bar{\sigma}_J^2) \xi^2}{1 - \bar{\sigma}_J^2 \xi^2}.
\end{equation}

In what is to follow, we use definitions of $\Gamma_J$ and $\eta_{\textnormal{esc}}$ above to define the corresponding values for the case of finite $N_L$ by making the replacement
\begin{equation}
    \xi = e^{-\zeta(L_r)/2} \rightarrow \prod_{n \in \textnormal{ring}} \sigma?{ph}_{J, n},
\end{equation}
where the product of the $\sigma?{ph}_{J, n}$ is taken over those corresponding to phantom channels coupled to the ring.

Next, for a pump producing a pulse with a Gaussian intensity distribution in the real waveguide input channel, and vacuum in all other ports, we can write the initial pump fields \emph{in the asymptotic-in basis} as
\begin{equation} \label{eq:pulseDefinition}
\begin{split}
    \alpha?{in}_{P,0}(k, 0) = \left( \frac{2}{\pi} \right)^{1/4} \sqrt{\frac{E_p \tau_p v_p}{\hbar \omega_p}} e^{-v_p^2 \tau_p^2 (k-k_0)^2} e^{-i(k-k_0) \mu_z}  
\end{split}
\end{equation}
and $\alpha_{P,n}^{in}(k,0) = 0$ for $n\neq 0$, such that the pump intensity has a standard deviation of $\tau_p$ before arriving at the ring system; $E_p$ denotes the energy of the pulse, with $\mu_z$ the location of the pulse center at $t=0$ and $k_0$ the central wavenumber. In what follows, we set the wavelength of the pump to be $\lambda_p = 1550$ nm, with an effective index of $n_{e} = 2.0$ and a group velocity of $v_p = 1.5\times 10^8$ m/s. Transforming into the local basis, we find that the initial field in the spatial mode $n$ is given by 
\begin{equation}
\begin{split}
    \alpha?{loc}_{P, n}(k,0) &= \sum_m \left[ \mathcal{L}^{\textnormal{in}-1}_{J,k} \right]_{m, n} \alpha_{P,m}?{in}(k,0) \\
    &= \left[ \mathcal{L}^{\textnormal{in}-1}_{J,k} \right]_{0, n} \alpha_{P,0}?{in}(k,0)
\end{split}
\end{equation}

We take the signal and idler resonances to correspond to the nearest neighbor resonances of the pump (see Fig. \ref{fig:ThreeResonances}). For simplicity we assume that the group velocity dispersion and the variation of the mode index $n_e$ over the frequency range of the resonances of interest is sufficiently small that we can approximate $v_i = v_p = v_s = v$, and take the central frequencies of the signal and idler fields, $\omega_s$ and $\omega_i$, to be given by
\begin{equation}
    \begin{split}
        \omega_s &= \omega_p + \frac{c}{n_{e} R_{e}} \\
        \omega_i &= \omega_p - \frac{c}{n_{e} R_{e}}
    \end{split}
\end{equation}
With this, we place a total of $N_L^R = 20$ loss channels coupled around the ring, with the ranges $R(J)$ defined to span a interval of $n_r\Gamma_J$ centered about $\omega_J$, with $n_r$ sufficiently large to capture the full evolution of the fields over a broad range along each resonance.

Solving for the evolution of the operators in the \emph{local basis} and transforming into the \emph{asymptotic-out} basis as in equation (\ref{eq:FinalTimeAsyOut}), and noting that the nonlinear interactions only mix $a_S$ with $a_S$ and $a_I^{\dagger}$ (and similarly $a_I$ with $a_I$ and $a_S^{\dagger}$), we can write the field operators at the final time $t_f$ as 
\begin{equation}
\begin{split}
    a_{S, n}?{out}(k, t_f) &= \sum_{n', k'} \left(V_{n,n'}^{SS}(k, k') a?{out}_{S, n'}(k', t_i) \right. \\
    & \qquad \qquad \left. + W_{n,n'}^{SI}(k, k') a^{\textnormal{out} \dagger}_{I, n'}(k', t_i) \right) \\
    a_{I, n}?{out}(k, t_f) &= \sum_{n', k'} \left(V_{n,n'}^{II}(k, k') a?{out}_{I, n'}(k', t_i) \right. \\
    & \qquad \qquad \left. + W_{n,n'}^{IS}(k, k') a^{\textnormal{out} \dagger}_{S, n'}(k', t_i) \right)
\end{split}
\end{equation}
with $W_{n,n'}^{JJ'}(k,k')$ and $V_{n,n'}^{JJ'}(k,k')$ corresponding to matrix elements of the full time evolution operator, $\mathcal{U}^{out}(t_i, t_f)$. We then solve for the second order moments of the generated fields as
\begin{equation}
\begin{split}
    &\langle a^{\textnormal{out} \dagger}_{J, n}(k, t_f) a^{out}_{J, n'}(k', t_f) \rangle \\ 
    & \qquad \qquad \qquad = \sum_{n'', k''} W^{JJ'*}_{n, n''}(k, k'') W^{JJ'}_{n', n''}(k', k'') \\
    &\langle a?{out}_{J, n}(k, t_f) a?{out}_{J', n'}(k', t_f) \rangle \\ 
    & \qquad \qquad \qquad = \sum_{n'', k''} V^{JJ}_{n, n''}(k, k'') W^{J'J}_{n', n''}(k', k'') \\ 
    &\langle a?{out}_{J, n}(k, t_f) a?{out}_{J, n'}(k', t_f) \rangle = 0 = \langle a^{\textnormal{out} \dagger}_{J, n}(k, t_f) a^{\textnormal{out} \dagger}_{J, n'}(k', t_f) \rangle
\end{split}
\end{equation}
where $J, J' \in \{S, I\}$ with $J' \neq J$. Note that the total number of photons generated in the resonance $J$ can be computed as
\begin{equation}
\begin{split}
    n?{tot}_J &= \sum_{n,k} \langle a^{\textnormal{out} \dagger}_{J, n}(k, t_f) a?{out}_{J, n}(k, t_f) \rangle \\
    &= \sum_{n, n', k,  k'} W^{JJ'*}_{n, n'}(k, k') W^{JJ'}_{n, n'}(k, k'),
\end{split}
\end{equation}
and since there is only a single SFWM interaction generating pairs in the signal and idler resonances, it follows that $n?{tot}_S = n?{tot}_I$.

\begin{figure}[t!]
    \centering
    \includegraphics[width=0.45\textwidth]{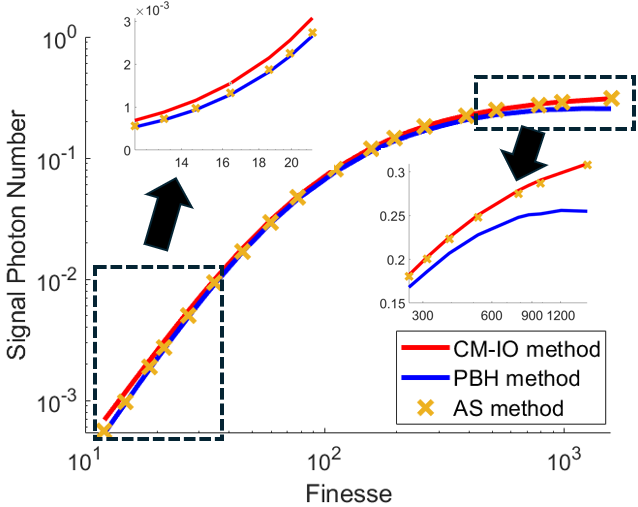}
    \caption{Number of generated signal photons per pulse, neglecting contributions from SPM and XPM. In all cases, the pulse energy and standard deviation are $E_p = 100$ pJ and $\tau_p = 70$ ps, with the escape efficiency of the ring fixed at $\eta_{\textnormal{esc}} = 0.75$ and $k_0 = k_p$.}
    \label{fig:FinesseSweep}
\end{figure}

In Fig. \ref{fig:FinesseSweep} we compare the predicted total number of signal photons, with the neglect of SPM and XPM, generated by an input pump pulse with $\tau_p=70$ ps and energy $E_p=100$ pJ using three separate methods; the proposed asymptotic scattering method (AS method), a coupled mode IO method (CM-IO method) \cite{Quesada:22, Helt:10, Vernon-PhysRevA.92.033840}, and a perturbative backwards Heisenberg method (PBH method) based on a previous asymptotic treatment \cite{Liscidini}, which has been used for the simulation of both  $\chi_2$ \cite{Liscidini} and $\chi_3$ processes \cite{Banic-PhysRevA.106.043707, Onodera:16}. Building on top of this, here we use a numerical implementation of the PBH, with the addition of a single phantom channel to take into account lost and broken pairs, as suggested earlier \cite{Banic-PhysRevA.106.043707}. Unlike what was done in that study \cite{Banic-PhysRevA.106.043707}, here we use a Gaussian pump wavefunction as outlined in T. Onodera et al. \cite{Onodera:16} to compare with the pulsed pump operation presented in this work. As expected, the proposed AS method matches well with the PBH method within the regime in which the total number of generated photons pairs is small ($n?{tot}_s \ll 1$) and thus the first order perturbative approximation is valid, but the methods begin to disagree at higher finesse when the expected number of pairs approach $n?{tot}_S \sim 1$. On the other hand, the AS method and CM-IO method agree at high finesse where the latter is expected to be valid, but disagree at low finesse.

\begin{figure}[t!]
    \centering
    \includegraphics[width=0.45\textwidth]{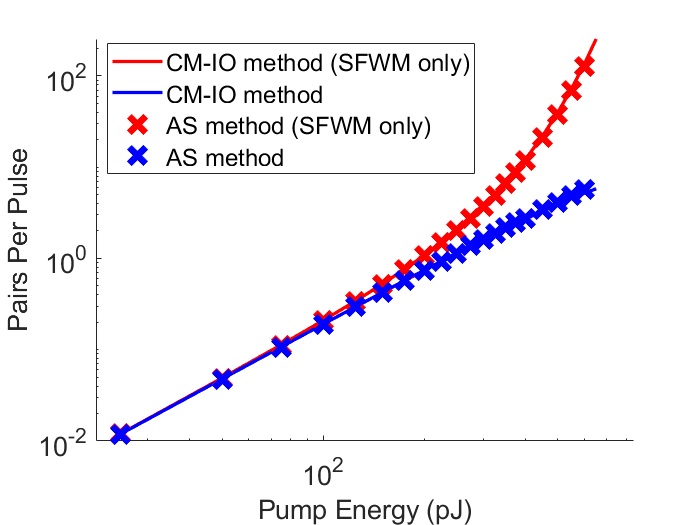}
    \caption{Number of signal photons per pulse as a function of pump power when including (blue) and neglecting (red) SPM and XPM. Here, the finesse of the ring is $\mathcal{F} = 780$ with $\tau_p = 70$ ps and $k_0 = k_p$.}
    \label{fig:PowerSweep}
\end{figure}

The agreement between the AS and CM-IO methods in high finesse regimes can be demonstrated far beyond the perturbative regime, even when $n?{tot}_s \gg 1$, as shown in Fig. \ref{fig:PowerSweep}. This is true even when including additional nonlinear processes such as SPM and XPM, and thus expands the regime of applicability for asymptotic field based methods compared to previous treatments \cite{Liscidini, Banic-PhysRevA.106.043707}.

\begin{figure}
    \centering
    \includegraphics[width=0.48\textwidth]{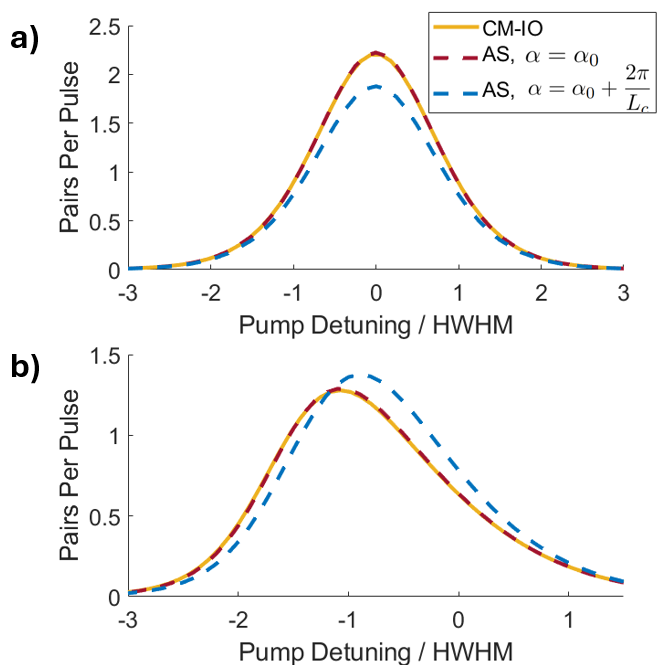}
    \caption{Expected number of generated photon pairs for a 300 pJ Gaussian pulse with $\tau_p = 140$ ps incident on a ring with finesse of $\mathcal{F} = 156$ both (a) neglecting and (b) including SPM and XPM. Dotted red line corresponds to the AS method with a weakly coupled ring ($\alpha_J L_c \ll \pi$), with the dotted blue line corresponding to a more strongly coupled system of equal finesse in which the ring and waveguide field oscillate once within the coupling region.}
    \label{fig:wigglePlot}
\end{figure}

\begin{figure}
    \centering
    \includegraphics[width=0.45\textwidth]{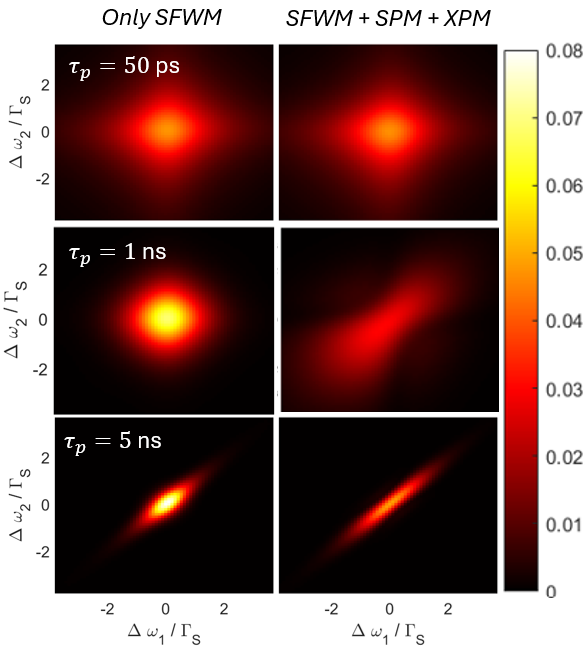}
    \caption{$|\bar{N}_{SS}|/n^{out}_s$ for input pulses with $\tau_p$ (see eq. \ref{eq:pulseDefinition}) of $50$ ps (top row), $1$ ns (middle row), and $5$ ns (bottom row), while including (right column) and neglecting (left column) contributions from SPM and XPM. Here, $\Gamma_S = 128$ MHz and corresponds to the half width at half maximum of the signal resonance.}
    \label{fig:Nmoments}
\end{figure}

We note, however, that the coupled mode approach is derived under the assumption of high finesse and weakly coupled rings. If instead we consider a more strongly coupled system in which $\alpha_J \rightarrow \alpha_J + \frac{2\pi}{L_c}$ such that the net single pass self- and cross-coupling parameters, $\bar{\sigma}_J$ and $\bar{\kappa}_J$ - and as a result the quality factor and finesse - of the system remains the same, then we begin to see differences between the AS and CM-IO methods as shown in Fig. (\ref{fig:wigglePlot}). This is due to the ability to resolve the position dependent variation in the field intensity within the coupling region when using the AS method. This increased mixing of the waveguide and resonator field can in many cases lead to a reduction in the generated number of photons, due to the more intense ring field being split between the ring and waveguide within the coupling region. But this also leads to a variation in the phase accumulated through SPM and XPM, leading to modifications of the field dynamics of the pump and generated signal and idler field. Such modifications can depend significantly on the quality factor of the coupled ring, as well as the pump energy and pulse shape, and consequently would be difficult to approximate with an effective decrease (increase) of the CM-IO nonlinear parameter. The ability to resolve the field variation in this way would be necessary in the the description of squeezing in linearly uncoupled dual-ring systems \cite{Tan:20, Menotti}, in which the pair generation is appreciable only within the coupling region shared by the two rings.

To analyze the field exiting the ring into the waveguide output channel, we construct the $N_k \times N_k$ matrices $\bar{N}_{J,J'}$ and $\bar{M}_{J,J'}$, defined as
\begin{equation} \label{eq:outputMoments}
\begin{split}
    \left[\bar{N}_{J, J'}\right]_{i,j} &= \langle a^{\textnormal{out} \dagger}_{J, 0}(k_i, t_f) a?{out}_{J', 0}(k_j, t_f) \rangle \\
    \left[\bar{M}_{J, J'}\right]_{i,j} &= \langle a?{out}_{J, 0}(k_i, t_f) a?{out}_{J', 0}(k_j, t_f) \rangle
\end{split}
\end{equation}
from which many important properties of the state can be derived. As an example, it is straightforward to see that the number of photons in the resonance $J$, exiting from the waveguide output, $n?{out}_J$, is simply given by
\begin{equation}
    n?{out}_J = \Tr \left[ \bar{N}_{JJ} \right].
\end{equation} 

Typically, the spectral correlations present in the output fields can be understood by appealing to the joint spectral amplitude (JSA) \cite{Quesada:22, Banic-PhysRevA.106.043707, Vernon-PhysRevA.92.033840}. However, such a quantity cannot easily be constructed through the $W^{JJ'}_{n,n'}(k,k')$ and $V^{JJ'}_{n,n'}(k,k')$ matrices describing the state at the final time $t_f$. Rather, one can visualize these spectral correlations through the $\bar{N}_{SS}$ distribution (eq. \ref{eq:outputMoments}) as shown in Fig. (\ref{fig:Nmoments}) for a pulse with $\tau_p$ = 50 ps, 1 ns, and 5 ns. Indeed, when $\tau_p$ is increased, the spectral bandwidth of the pump pulse decreases, leading to a narrower $\bar{N}_{SS}$. The inclusion of SPM and XPM then induces a time dependent change in the index of refraction for the pump and generated fields, broadening the $\bar{N}_{SS}$ distribution.

\begin{figure}[t!]
    \centering
    \includegraphics[width=0.45\textwidth]{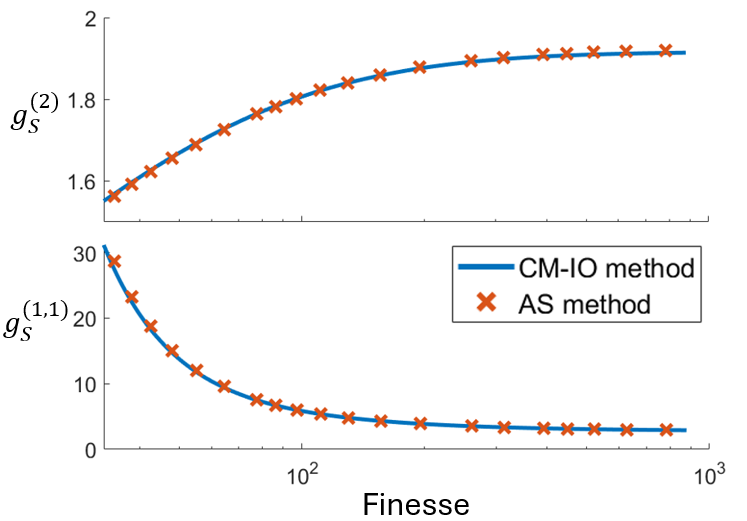}
    \caption{$g^{(2)}_{S}$ and $g^{(1,1)}_{SI}$ correlation values as a function of the finesse of the ring system. Pulse is fixed with $E_p = 200$ pJ and $\tau_p = 70$ ps, and an escape efficiency of the ring of $\eta_{\textnormal{esc}}=0.75$.}
    \label{fig:gPlots}
\end{figure}

In addition to the number of generated pairs, we can also utilize the $\bar{N}_{JJ'}$ and $\bar{M}_{JJ'}$ to determine the $n^{th}$ order correlation functions of the output fields. In particular, we will consider the broadband second order correlation and cross-correlation values, $g_{J}^{(2)}$ and $g_{JJ'}^{(1,1)}$ \cite{Christ_2011},
\begin{equation} 
\begin{split}
    g^{(2)}_J &= \frac{\int dt_1 dt_2 \langle \hat{E}_J^{(-)}(t_1) \hat{E}_J^{(-)}(t_2) \hat{E}_J^{(+)}(t_2) \hat{E}_J^{(+)}(t_1) \rangle}{\int dt_1 dt_2 \langle \hat{E}_J^{(-)}(t_1) \hat{E}_J^{(+)}(t_1) \rangle \langle \hat{E}_J^{(-)}(t_2) \hat{E}_J^{(+)}(t_2) \rangle} \\
    g^{(1,1)}_{JJ'} &= \frac{\int dt_1 dt_2 \langle \hat{E}_J^{(-)}(t_1) \hat{E}_J^{(+)}(t_1) \hat{E}_{J'}^{(-)}(t_2) \hat{E}_{J'}^{(+)}(t_2) \rangle}{\int dt_1 dt_2 \langle \hat{E}_J^{(-)}(t_1) \hat{E}_J^{(+)}(t_1) \rangle \langle \hat{E}_{J'}^{(-)}(t_2) \hat{E}_{J'}^{(+)}(t_2) \rangle}, \\
\end{split}
\end{equation}
where the integration window is taken to be much longer than the pulse duration. Restricting the fields to the output waveguide, using the form of the output fields and evaluating each of the integrals over time, we find \cite{Quesada:22, Christ_2011} 
\begin{widetext}
\begin{equation}
    \begin{split}
        g^{(2)}_J &= \frac{\sum_k \sum_{k'} \langle a^{\textnormal{out} \dagger}_{J, 0}(k, t_f) a^{\textnormal{out} \dagger}_{J, 0}(k', t_f) a?{out}_{J, 0}(k', t_f) a?{out}_{J, 0}(k, t_f) \rangle }{\left[ \sum_k \langle a^{\textnormal{out} \dagger}_{J,0}(k, t_f) a?{out}_{J,0}(k, t_f) \rangle \right]^2} = \frac{\Tr\left[ \bar{N}_{JJ}^2 \right] + \Tr\left[ \bar{N}_{JJ} \right]^2}{\Tr\left[ \bar{N}_{JJ} \right]^2} \\
        g^{(1,1)}_{JJ'} &= \frac{\sum_k \sum_{k'} \langle a^{\textnormal{out} \dagger}_{J, 0}(k, t_f) a?{out}_{J, 0}(k, t_f) a^{\textnormal{out} \dagger}_{J', 0}(k', t_f) a?{out}_{J', 0}(k', t_f) \rangle }{\left[ \sum_k \langle a^{\textnormal{out} \dagger}_{J,0}(k, t_f) a?{out}_{J,0}(k, t_f) \rangle \right] \left[ \sum_{k'} \langle a^{\textnormal{out} \dagger}_{J',0}(k', t_f) a?{out}_{J',0}(k', t_f) \rangle \right]} = \frac{\Tr\left[ \bar{M}_{JJ'}\bar{M}^{\dagger}_{JJ'} \right] + \Tr\left[ \bar{N}_{JJ} \right]\Tr\left[ \bar{N}_{J'J'} \right]}{\Tr\left[ \bar{N}_{JJ} \right]\Tr\left[ \bar{N}_{J'J'} \right]} \\
    \end{split}
\end{equation}
\end{widetext}
Fig. \ref{fig:gPlots} shows the correlation values for a pulse with $\tau_p = 70$ ps and $E_p = 200$ pJ, as a function of finesse with a fixed escape efficiency and weak coupling. The AS method matches well with the predictions from a standard coupled-mode treatment. 

We note that while the values $g_J^{(2)}$ and $g_{JJ'}^{(1,1)}$ are typically related to an effective mode number (or Schmidt number) of the output state, such a treatment is not possible here as the output after tracing out over the phantom channels (or any filtering) can no longer be described as a pure squeezed state. Rather, one could define matrices $N_{JJ}$ and $M_{JJ'}$ similarly to what was done in equation (\ref{eq:outputMoments}), but including all modes of the system, and define a Schmidt mode basis including contributions from the asymptotic-out operators for each phantom channel and the corresponding unitary transformations $U_J$ such that $N_{JJ} = U_J^* D_N U_J^T$ and $M_{JJ'} = U_J D_M U_{J'}^T$ with $D_N$ and $D_M$ diagonal. By restricting our attention only to those asymptotic-out modes of the real waveguide output channel, a decomposition of the $\bar{N}_{JJ}$ and $\bar{M}_{JJ'}$ matrices can still be performed as $\bar{N}_{JJ} = F_J^* \bar{D}_N F_J^T$ and $\bar{N}_{JJ'} = G_J^* \bar{D}_M G_{J'}^T$ for $\bar{D}_N$ and $\bar{D}_M$ diagonal. But it is no longer guaranteed that $F_J = G_J$, as is the case with a general pure squeezed state. Consequently, a Schmidt mode basis of the output field is not well defined.

\begin{figure}
    \centering
    \includegraphics[width=0.45\textwidth]{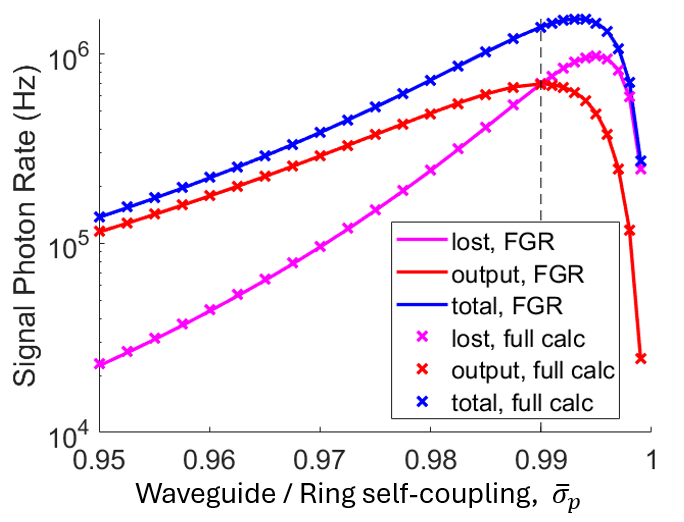}
    \caption{Rate of lost (magenta), output (red), and total signal photon generation (blue) for a CW pump with $P_p = 10$ mW, $k_0 = k_p$, and neglecting SPM and XPM. Solid lines correspond to a Fermi's Golden Rule (FGR) rate calculation, with the crosses denoting a full non-perturbative calculation as presented here. Dashed line represents the critical coupling of the ring and the waveguide. }
    \label{fig:CWrates}
\end{figure}

To end this section, we also consider a quasi-CW pump scenario, in which we take 
\begin{equation}
    a?{in}_{P, n}(k, 0) = \begin{cases} \sqrt{\frac{2\pi P_p}{\hbar \omega_p v_p}} \delta (k - k_0)  & \textit{for } n=0 \\
    0 & \textit{otherwise}
    \end{cases}
\end{equation}
and the initial state of the signal and idler fields to be the vacuum; we then allow the system to evolve into a steady state. Here $P_p$ identifies the power of the input pump in the real waveguide input, with $k_0$ the wavenumber. From this we can calculate a rate of generated pairs which, in the absence of SPM and XPM, matches well with a Fermi's Golden rule calculation based on previous treatments using the asymptotic field expansion \cite{Banic-PhysRevA.106.043707} as shown in Fig. \ref{fig:CWrates}.

Finally, we consider the quadrature variance of the resulting squeezed output state. This can be identified experimentally by, for example, mixing the signal and idler output with a bi-chromatic local oscillator at a 50/50 beam splitter, and measuring the difference photo-current \cite{Vaidya-doi:10.1126/sciadv.aba9186}. For $\phi = \phi_S + \phi_I$, with $\phi_J$ being the relative phase of the local oscillator beam stimulating the frequency $\omega_J$, this leads to a relative variance of
\begin{equation}
\begin{split}
    V(\omega; \phi) &= 1 + \tilde{N}(\omega, \omega) + \tilde{N}(-\omega, -\omega) \\
    & \qquad \qquad + 2 \Re \left\{ \tilde{M}(\omega, -\omega) e^{-i\phi} \right\},
\end{split}
\end{equation}
where,
\begin{equation} \label{eq:QuadratureVarience}
    \begin{split}
        \tilde{N}(\omega, \omega') &= \frac{1}{2} \left[ \bar{N}_{SS} \left( k_S + \frac{\omega}{v_S}, k_S + \frac{\omega'}{v_S} \right) \right. \\
        &\left. \qquad \qquad + \bar{N}_{II} \left( k_I + \frac{\omega}{v_I}, k_I + \frac{\omega'}{v_I} \right) \right] \\
        \tilde{M}(\omega, \omega') &= \frac{1}{2} \left[ \bar{M}_{SI} \left( k_S + \frac{\omega}{v_S}, k_I + \frac{\omega'}{v_I} \right) \right. \\
        &\left. \qquad \qquad + \bar{M}_{IS} \left( k_I + \frac{\omega}{v_I}, k_S + \frac{\omega'}{v_S} \right) \right]
    \end{split}
\end{equation}

\begin{figure}
    \centering
    \includegraphics[width=0.45\textwidth]{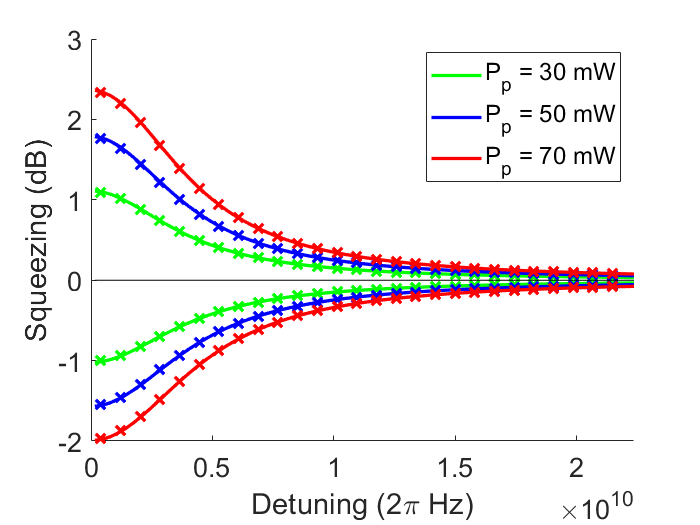}
    \caption{Squeezing spectrum for a CW pump with finesse of $\mathcal{F} = 780$ and escape efficiency of $\eta_{\textnormal{esc}}=0.75$. Solid lines correspond to an analytic calculation involving the CM-IO method, where as the crosses correspond to the AS method.}
    \label{fig:SqueezingSpectrum}
\end{figure}

Fig. \ref{fig:SqueezingSpectrum} shows the squeezing and anti-squeezing spectrum resulting from the maximization / minimization of equation (\ref{eq:QuadratureVarience}) with respect to the phase $\phi$, for a number of pump powers. This can be seen to match well with analytic expressions for the squeezing derived from a coupled mode treatment of a CW input \cite{Vaidya-doi:10.1126/sciadv.aba9186}.

\section{Conclusion} \label{sec:Conclusion}
We have developed a method for modeling photon pair generation through SFWM processes in a microring system beyond the perturbative regime, utilizing an asymptotic field description of the field modes. This has been shown to match well with a standard coupled-mode calculation for non-degenerate, single-pump squeezing scheme in the case of high finesse and weakly coupled resonators through a number of metrics, such as pair generation rate, squeezing spectrum, and second order correlation values of the output fields. Additionally, the proposed \emph{non-perturbative} asymptotic field based method have been shown to match well with previous \emph{perturbative} asymptotic field based methods in the regime of low pair generation rate. As such, this asymptotic scattering method expands upon the regime of applicability of both the coupled mode calculations and perturbative backwards Heisenberg calculations, being applicable for low finesse resonators and high pair generation rates, while including additional nonlinear processes such as SPM and XPM and allowing for a local treatment of the field variation in the coupling region.

Owing to a larger mode dimensionality introduced by the phantom channels, the asymptotic scattering method developed here does lead to longer run times for numerical calculations. But in many regimes of practical interest, such as the ring resonators considered here, the total number of phantom channels can be kept modest.

Furthermore, our method is readily generalizable to more complicated structures, needing only the derivation of the asymptotic field distributions in any new system. This paves the way for future studies of photonic molecules including multiple coupled rings where linear effects such as the splitting of shared resonances come ``prepackaged" in the asymptotic modes. This can be done while also providing the flexibility of a general coupling description between optical elements along a finite region, and generalizations of the approach allowing for a local treatment of effects such as higher-order spatial mode coupling in the resonator bends and back-scattering. Studies of linearly uncoupled resonators, in which the energy transfer between elements happens only from nonlinear interactions between the fields, can be readily implemented in our approach, while they would be much more difficult to undertake using the linear coupling assumptions of many coupled mode schemes.

\section*{Acknowledgements} \label{sec:acknowledgements}
MS acknowledges support from the Mitacs Accelerate program, and the Ontario Graduate Scholarship program. MS also Acknowledges support from Xanadu Quantum Technologies for use of proprietary code in coupled mode calculations. ML acknowledges the support by PNRR MUR project PE0000023-NQSTI. JES acknowledges support from the Natural Sciences and Engineering Research Council of Canada.

\bibliographystyle{unsrt}
\bibliography{References}

\appendix

\section{Classical Field Solution For a Lossless Ring} \label{sec:appendix1}
We start by considering the coupled system of equations in equation (\ref{eq:EOMofLinearFields}) within the coupling region ($0 < z \leq L_c$). In particular, we are interested in solutions in which the operators $\psi_J(z,t)$ and $\phi_J(z,t)$ are oscillating at a frequency $\omega_k$ near $\omega_J$. Therefore, after replacing the operators with their expectation values as in (\ref{eq:classicalApproximation}), we search for solutions of the form
\begin{equation}
    \begin{split}
        \langle \psi_J(z, t; k) \rangle &= \langle \psi_J(0, 0; k) \rangle \tilde{f}?{wg}_{J}(z) e^{-i\omega_k t} \\
        \langle \phi_J(z, t; k) \rangle &= \langle \psi_J(0, 0; k) \rangle \tilde{f}?{rr}_{J}(z) e^{-i\omega_k t}
    \end{split}
\end{equation}
Writing $\bar{f}?{wg}_{J}(z) = \tilde{f}?{wg}_{J}(z) e^{i \Delta \beta_J z/2}$ and $\bar{f}?{rr}_{J}(z) = \tilde{f}?{rr}_{J}(z) e^{-i \Delta \beta_J z/2}$, we put the form of the solution above in equation (\ref{eq:EOMofLinearFields}) and rearrange to find
\begin{equation} \label{eq:simplifiedEquationOfMotion}
\begin{split}
    \frac{d}{dz} \begin{pmatrix}
        \bar{f}?{wg}_{J}(z) \\ \bar{f}?{rr}_{J}(z)
    \end{pmatrix} &= i
    \begin{pmatrix}
        \mu_{J,1}(k) & -\omega_c/v_J \\
        -\omega_c/u_J & \mu_{J,2}(k)
    \end{pmatrix} \begin{pmatrix}
        \bar{f}?{wg}_{J}(z) \\ \bar{f}?{rr}_{J}(z)
    \end{pmatrix} \\
    &= iM_J(k) \begin{pmatrix}
        \bar{f}?{wg}_{J}(z) \\ \bar{f}?{rr}_{J}(z)
    \end{pmatrix},
\end{split}
\end{equation}
where we have introduced the functions $\mu_{J,1}(k)$ and $\mu_{J,2}(k)$ as 
\begin{equation}
    \begin{split}
        \mu_{J,1}(k) &= \frac{\omega_k - \omega_J}{v_J} + \frac{\Delta \beta_J}{2} \\
        \mu_{J,2}(k) &= \frac{\omega_k - \omega_J}{u_J} - \frac{\Delta \beta_J}{2}
    \end{split}
\end{equation}

Defining $\mu_{J,\pm}(k) = (\mu_{J,1}(k) \pm \mu_{J,2} (k))/2$ and the wavenumber $\alpha_k = \sqrt{\mu_{J,-}^2(k) + \alpha_{J,0}^2}$ for $\alpha_{J,0} = \frac{\omega_c}{\sqrt{v_J u_J}}$, we expand $M_J(k)$ as
\begin{equation}
    M_J(k) = P_J(k) \begin{pmatrix}
        \lambda^+_k & 0 \\ 0 & \lambda^-_k
    \end{pmatrix} P_J^{-1}(k),
\end{equation}
where the matrix $P_J(k)$ is given as
\begin{equation}
    P_J(k) = \begin{pmatrix}
        \omega_c/v_J & -\left(\mu_{J,-}(k) - \alpha_k \right) \\
        \mu_{J,-}(k) - \alpha_k & \omega_c/ u_J
    \end{pmatrix},
\end{equation}
and $\lambda^{\pm}_k = \mu_{J,+}(k) \pm \alpha_k$.
Then we can write the solutions to equation (\ref{eq:simplifiedEquationOfMotion}) as 
\begin{widetext}
\begin{equation}
    \begin{pmatrix}
        \bar{f}?{wg}_{J}(z) \\ \bar{f}?{rr}_{J}(z)
    \end{pmatrix} = P_J(k) \begin{pmatrix}
        e^{i \lambda^+_k z} & 0 \\ 0 & e^{-i \lambda^-_k z}
    \end{pmatrix} P_J^{-1}(k) \begin{pmatrix}
        \bar{f}?{wg}_{J}(0) \\ \bar{f}?{rr}_{J}(0)
    \end{pmatrix} = 
    \begin{pmatrix}
        \tilde{\sigma}_J(z; k) & -i \sqrt{\frac{u_J}{v_J}} \tilde{\kappa}_J(z; k) \\ -i \sqrt{\frac{v_J}{u_J}} \tilde{\kappa}_J(z; k) & \tilde{\sigma}_J^*(z; k)
    \end{pmatrix}
    \begin{pmatrix}
        \bar{f}?{wg}_{J}(0) \\ \bar{f}?{rr}_{J}(0)
    \end{pmatrix} e^{i \Delta k l_J(z)}
\end{equation}
\end{widetext}
for the functions $\sigma_J(z; k)$ and $\tilde{\kappa}_J(z; k)$ defined as
\begin{equation}
    \begin{split}
        \tilde{\sigma}_J(z; k) &= \cos \left( \alpha_k z \right) + i\gamma_J(k) \sin \left( \alpha_k z \right), \\
        \tilde{\kappa}_J(z;k) &= \sqrt{1 - \gamma_J(k)^2} \sin \left( \alpha_k z \right).
    \end{split}
\end{equation}
Transforming back to the $\tilde{f}^{wg}_{J}(z)$ and $\tilde{f}^{rr}_{J}(z)$ basis results in a solution of the form 
\begin{equation}
\begin{split}
    \begin{pmatrix}
        \tilde{f}?{wg}_{J}(z) \\ \tilde{f}?{rr}_{J}(z)
    \end{pmatrix} &= 
    \begin{pmatrix}
        \sigma_J(z; k) & -i \sqrt{\frac{u_J}{v_J}} \kappa_J(z; k) \\ -i \sqrt{\frac{v_J}{u_J}} \kappa_J^*(z; k) & \sigma_J^*(z; k)
    \end{pmatrix} \\
    & \qquad \qquad \times
    \begin{pmatrix}
        \tilde{f}?{wg}_{J}(0) \\ \tilde{f}?{rr}_{J}(0)
    \end{pmatrix}  e^{i \mu_{J,+}(k) z},
\end{split}
\end{equation}
for $\sigma_J(z;k) = \tilde{\sigma}_J(z;k) e^{-i \Delta \beta_J z/2}$ and $\kappa_J(z; k) = \tilde{\kappa}_J(z; k) e^{-i \Delta \beta_J z/2}$ as in equation (\ref{eq:sigmaKappaDef}). 

On the other hand, for $\textbf{r}$ in the ring outside the coupling region ($L_c < z \leq L_r$), the solution for the field amplitude $\tilde{\beta}?{rr}_J(z)$ can be found simply as
\begin{equation}
    \tilde{f}?{rr}_J(z) = \tilde{f}?{rr}_J(L_c) e^{i (\omega_k - \omega_J) (z - L_c) / u_J}.
\end{equation}
Expanding $\omega$ in terms of $k$ as in equation (\ref{eq:waveguideDispersionRelation}), it follows that 
\begin{equation}
    \begin{split}
        \mu_{J,+}(k) &= \frac{1}{2} \left( 1 + \frac{v_J}{u_J} \right) \Delta k_J, \\
        \frac{\omega_k - \omega_J}{u_J} &= \frac{v_J}{u_J} \Delta k_J.
    \end{split}
\end{equation}
To construct the full steady state solution of the classical lossless system then only requires specifying the field amplitude at the waveguide input. As such, the particular solution presented in equation (\ref{eq:sampleClassicalSolution}) comes by setting $\tilde{f}?{wg}_J(0) = 1$.

\section{Local Basis Field Distributions} \label{sec:appendix2}
For completeness, we outline a sample local basis field distribution associated with a non-zero support localized within the ring and waveguide coupling region. Due to the field oscillating between the waveguide and the ring, it is not possible in general to construct a basis of modes within the coupling region that is disjoint from all other local basis field distributions. However, as outlined in Fig. (\ref{fig:LocalBasisDiagrams}), it is possible to construct pairs of modes which have some non-zero overlap with each other, but are disjoint from every other mode. 

In particular, suppose we label the loss channels such that the $n^{th}$ and $(n+1)^{th}$ loss channels are adjacent and coupled to the waveguide, while the $m^{th}$ and $(m+1)^{th}$ loss channels correspond to adjacent channels along the ring with $\bar{z}_m = \bar{z}_n$ and $\bar{z}_{m+1} = \bar{z}_{n+1}$. Taking a linear combination of the corresponding asymptotic-in field distributions, one could construct $D?{loc}_{J,k,n}(\textbf{r})$ and $D?{loc}_{J,k,m}(\textbf{r})$ as

\begin{widetext}
\begin{equation}
    \begin{split}
        D?{loc}_{J,k,n}(\textbf{r}) &= D?{asy-in}_{J,k,n}(\textbf{r}) - \left[ f^{n,n+1}_{J,k} D?{asy-in}_{J,k,n+1}(\textbf{r}) + f^{n,m+1}_{J,k} D?{asy-in}_{J,k,m+1}(\textbf{r}) \right] \\
        & = \begin{cases}
            \textbf{d}?{ph}_{J,n}(x,y) e^{ik_{[n]} z} & \textit{in the } n^{th} \textit{ loss channel and } z<0 \\
            - f^{n,n+1}_{J, k} \textbf{d}?{ph}_{J,n+1}(x,y) e^{ik_{[n+1]} z} & \textit{in the } (n+1)^{th} \textit{ loss channel and } z<0 \\
            -f^{n,m+1}_{J,k} \textbf{d}?{ph}_{J,m+1}(x,y) e^{ik_{[m+1]} z} & \textit{in the } (m+1)^{th} \textit{ loss channel and } z<0 \\
            -i\kappa?{ph}_{J,n} \textbf{d}?{asy}_{J,0}(x,y) \sigma_J(z-\bar{z}_n;k) & \\
            \qquad \qquad \times e^{i \Delta k_J (l_J(z) - l_J(\bar{z}_n))} e^{i k_J (z - \bar{z}_n)} & \textit{in the waveguide and } \bar{z}_n < z < \bar{z}_{n+1} \\
            -\sqrt{\frac{v_J}{u_J}} \kappa?{ph}_{J,n} \textbf{d}?{asy}_{J,0}(x,y) \kappa_J^*(z - \bar{z}_n;k) & \\ 
            \qquad \qquad \times e^{i \Delta k_J (l_J(z) - l_J(\bar{z}_n))} e^{i k_J' (z - \bar{z}_n)}  & \textit{in the ring and } \bar{z}_n < z < \bar{z}_{n+1} \\
            \sigma?{ph}_{J,n}\textbf{d}?{ph}_{J,n}(x,y) e^{ik_{[n]} z} & \textit{in the } n^{th} \textit{ loss channel and } z>0 \\
            -\frac{1}{\sigma?{ph}_{J,n+1}} f^{n,n+1}_{J, k} \textbf{d}?{ph}_{J,n+1}(x,y) e^{ik_{[n+1]} z} & \textit{in the } (n+1)^{th} \textit{ loss channel and } z>0 \\
            -\frac{1}{\sigma?{ph}_{J,m+1}} f^{n,m+1}_{J,k} \textbf{d}?{ph}_{J,m+1}(x,y) e^{i(k_{[m+1]} - k_J) z} & \textit{in the } (m+1)^{th} \textit{ loss channel and } z>0 \\
            0 & \textit{otherwise}
        \end{cases}
    \end{split}
\end{equation}

\begin{equation}
    \begin{split}
        D?{loc}_{J,k,m}(\textbf{r}) &= D?{asy-in}_{J,k,m}(\textbf{r}) - \left[ f^{m,n+1}_{J,k} D?{asy-in}_{J,k,n+1}(\textbf{r}) + f^{m,m+1}_{J,k} D?{asy-in}_{J,k,m+1}(\textbf{r}) \right] \\
        & = \begin{cases}
            \textbf{d}?{ph}_{J,m}(x,y) e^{ik_{[m]} z} & \textit{in the } m^{th} \textit{ loss channel and } z<0 \\
            - f^{m,n+1}_{J, k} \textbf{d}?{ph}_{J,n+1}(x,y) e^{ik_{[n+1]} z} & \textit{in the } (n+1)^{th} \textit{ loss channel and } z<0 \\
            -f^{m,m+1}_{J,k} \textbf{d}?{ph}_{J,m+1}(x,y) e^{ik_{[m+1]} z} & \textit{in the } (m+1)^{th} \textit{ loss channel and } z<0 \\
            -\sqrt{\frac{u_J}{v_J}} \kappa?{ph}_{J, m} \textbf{d}?{asy}_{J,0}(x,y) \kappa_J(z-\bar{z}_n;k) & \\
            \qquad \qquad \times e^{i \Delta k_J (l_J(z) - l_J(\bar{z}_m))} e^{i k_J (z - \bar{z}_m)} & \textit{in the waveguide and } \bar{z}_m < z < \bar{z}_{m+1} \\
            -i \kappa?{ph}_{J,m} \textbf{d}?{asy}_{J,0}(x,y) \sigma_J^*(z - \bar{z}_m;k) & \\
            \qquad \qquad \times e^{i \Delta k_J (l_J(z) - l_J(\bar{z}_m))} e^{i k_J' (z - \bar{z}_m)}  & \textit{in the ring and } \bar{z}_m < z < \bar{z}_{m+1} \\
            \sigma?{ph}_{J,m}\textbf{d}?{ph}_{J,m}(x,y) e^{ik_{[m]} z} & \textit{in the } m^{th} \textit{ loss channel and } z>0 \\
            -\frac{1}{\sigma?{ph}_{J,n+1}} f^{m,n+1}_{J, k} \textbf{d}?{ph}_{J,n+1}(x,y) e^{ik_{[n+1]} z} & \textit{in the } (n+1)^{th} \textit{ loss channel and } z>0 \\
            -\frac{1}{\sigma?{ph}_{J,m+1}} f^{m,m+1}_{J,k} \textbf{d}?{ph}_{J,m+1}(x,y) e^{i(k_{[m+1]} - k_J) z} & \textit{in the } (m+1)^{th} \textit{ loss channel and } z>0 \\
            0 & \textit{otherwise}
        \end{cases}
    \end{split}
\end{equation}
\end{widetext}
where the coefficients for $D?{loc}_{J,k,n}(\textbf{r})$ are given by
\begin{equation}
    \begin{split}
        f^{n,n+1}_{J,k} &= \sigma?{ph}_{J,n+1} \frac{\kappa?{ph}_{J,n}}{\kappa?{ph}_{J,n+1}} \sigma_J(\bar{z}_{n+1} - \bar{z}_{n}; k) \\
        & \qquad \times e^{i\Delta k_J (l_J(\bar{z}_{n+1}) - l_J(\bar{z}_n))} e^{i k_J(\bar{z}_{n+1} - \bar{z}_{n})} \\
        f^{n, m+1}_{J,k} &= -i \sqrt{\frac{v_J}{u_J}} \sigma?{ph}_{J,m+1} \frac{\kappa?{ph}_{J,n}}{\kappa?{ph}_{J,m+1}} \kappa_J(\bar{z}_{n+1} - \bar{z}_{n}; k) \\
        & \qquad \times e^{i\Delta k_J (l_J(\bar{z}_{n+1}) - l_J(\bar{z}_n))} e^{i k_J(\bar{z}_{n+1} - \bar{z}_{n})} \\
    \end{split}
\end{equation}
and the coefficients for $D?{loc}_{J,k,m}(\textbf{r})$ being
\begin{equation}
    \begin{split}
        f^{m,m+1}_{J,k} &= \sigma?{ph}_{J,m+1} \frac{\kappa?{ph}_{J,m}}{\kappa?{ph}_{J,m+1}} \sigma_J^*(\bar{z}_{m+1} - \bar{z}_{m}; k) \\
        & \qquad \times e^{i\Delta k_J (l_J(\bar{z}_{m+1}) - l_J(\bar{z}_m))} e^{i k_J(\bar{z}_{m+1} - \bar{z}_{m})} \\
        f^{m, n+1}_{J,k} &= -i \sqrt{\frac{u_J}{v_J}} \sigma?{ph}_{J,n+1} \frac{\kappa?{ph}_{J,m}}{\kappa?{ph}_{J,n+1}} \kappa_J^*(\bar{z}_{m+1} - \bar{z}_{m}; k) \\
        & \qquad \times e^{i\Delta k_J (l_J(\bar{z}_{m+1}) - l_J(\bar{z}_m))} e^{i k_J(\bar{z}_{m+1} - \bar{z}_{m})} \\
    \end{split}
\end{equation}
Note that here the function $D?{loc}_{J, k, n}(\textbf{r})$ corresponds schematically to the diagram in Fig. \ref{fig:LocalBasisDiagrams}b, where as $D?{loc}_{J, k, m}(\textbf{r})$ corresponds schematically to Fig. \ref{fig:LocalBasisDiagrams}c.

\section{Nonlinear Short Time Evolution} \label{sec:appendix3}
    Here we derive the solution to equation (\ref{eq:CompactEOM}) as presented in equation (\ref{eq:FullEvoSolution}). Consider a time interval $\left[ t_i, t_f \right]$ short enough that each of the matrices $\mathcal{A}?{NL}_i(t)$ are approximately constant. In this case we can then suppress the time argument and take 
    \begin{equation}
        \mathcal{A}?{NL}_i(t) \rightarrow \mathcal{A}?{NL}_i,
    \end{equation}
    This allows us to write the nonlinear part of equation (\ref{eq:CompactEOM}) as
    \begin{equation} \label{eq:shortTimeEOM}
        \frac{\partial}{\partial t} \textbf{a}_i?{loc}(t) = i\mathcal{A}?{NL}_i \sum_{j=1}^{N_k} \textbf{a}_j?{loc}(t),
    \end{equation}
    for $t \in \left[ t_i, t_f \right]$. Letting $\tilde{\textbf{a}}(t) = \sum_{j=1}^{N_k} \textbf{a}_j?{loc}(t)$, we can then sum over the $i$ index in equation (\ref{eq:shortTimeEOM}) to write
    \begin{equation}
        \frac{\partial}{\partial t} \tilde{\textbf{a}}(t) = i \left[ \sum_i \mathcal{A}_i?{NL} \right] \tilde{\textbf{a}}(t) = i \tilde{\mathcal{A}}?{NL} \tilde{\textbf{a}}(t),
    \end{equation} 
    which has a solution
    \begin{equation}
        \tilde{\textbf{a}}(t) = e^{i (t-t_i) \tilde{\mathcal{A}}?{NL} } \tilde{\textbf{a}}(t_i).
    \end{equation}
    Using this in equation (\ref{eq:shortTimeEOM}) results in
    \begin{equation}
        \frac{\partial}{\partial t} \textbf{a}_i?{loc}(t) = i\mathcal{A}?{NL}_i e^{i (t-t_i) \tilde{\mathcal{A}}?{NL} } \sum_{j=1}^{N_k} \textbf{a}_j?{loc}(t_i),
    \end{equation}
    leading to the solution to $a?{loc}_i(t)$ for $t \in \left[t_i, t_f \right]$ given by
    \begin{equation}
    \begin{split}
        \textbf{a}?{loc}_i(t) &= \textbf{a}?{loc}_i(t_i) \\
        & + i\mathcal{A}?{NL}_i \left[ \sum_{n=1}^{\infty} \frac{(t - t_i)^{n+1}}{(n+1)!} \left( i \tilde{\mathcal{A}}?{NL} \right)^n \right]  \sum_{j=1}^{N_k} \textbf{a}_j?{loc}(t_i),
    \end{split}
    \end{equation}
    Taking $t_i = t$ and $t_f = t+\Delta t$, and reintroducing the time dependence of the $\mathcal{A}?{NL}_i(t)$ matrices, we end up with
    \begin{equation}
        \textbf{a}?{loc}_i(t+\Delta t) = \textbf{a}?{loc}_i(t) + U?{NL}_i(t, t+\Delta t) \sum_{j=1}^{N_k} \textbf{a}_j?{loc}(t_i),
    \end{equation}
    with the $U?{NL}_i(t, t+\Delta t)$ being given as in equation (\ref{eq:NLevoMats}).

\end{document}